\documentclass[preprint,12pt]{elsarticle}

\usepackage{subcaption}
\usepackage{multirow}

\usepackage{amssymb} 
\usepackage{amsthm} 
\usepackage{amsmath} 
\usepackage{amsfonts}
\usepackage{booktabs}
\usepackage{comment}
\usepackage{url}
\usepackage{here}
\usepackage[subrefformat=parens]{subcaption}
\usepackage{bm}
\usepackage{siunitx} 
\usepackage{afterpage} 
\usepackage{algorithmic} 
\usepackage{algorithm} 
\usepackage{afterpage} 

\journal{Journal of  Space Safety Engineering}

\begin{document}

\begin{frontmatter}



\title{Attitude Estimation from Photometric Data using Gaussian Process Regression}


\author[KU]{Ryui Hara}
\author[KU]{Yasuhiro Yoshimura}
\author[KU]{Toshiya Hanada}

\affiliation[KU]{organization={Department of Aeronautics and Astronautics, Kyushu Univerisity},
            addressline={744 Motooka, Nishi-ku}, 
            city={Fukuoka},
            postcode={819-0395}, 
            country={Japan}}

\begin{abstract}
The rapid growth of resident space objects in Earth's orbit has intensified the need for advanced space situational awareness and space domain awareness to manage satellite traffic and prevent collisions. Attitude estimation is critical for accurate state propagation, as non-gravitational forces like solar radiation pressure and atmospheric drag depend on the object's attitude. This study explores using light curves, 
time variation of an object's brightness
, to estimate a space object's attitude. Light curve inversion, traditionally used in astronomy, faces challenges when applied to resident space objects due to their non-convex shapes and specular reflections. Conventional methods for attitude estimation often assume known shape and surface parameters, which are usually unknown for space debris generated by a collision or breakup. To address this issue, this study proposes the estimation method combining Gaussian process regression with the unscented Kalman filter. This study uses Gaussian process regression for a non-parametric observation model, enhancing robustness against unknown surface parameters. Numerical examples consider a box-wing object in a geosynchronous orbit and demonstrate that the proposed method has better estimation accuracy than a conventional unscented Kalman filter. The numerical simulation results also represent the attitude estimation robust against uncertainties in surface properties, contributing to practical scenarios in space situational awareness and space domain awareness where the object parameters are unknown. 
\end{abstract}



\begin{keyword}
Attitude estimation \sep Gaussian process regression \sep Unscented Kalman filter


\end{keyword}

\end{frontmatter}


\section{Introduction}
Recent advances in large constellations drastically increase the population of resident space objects (RSOs) around the Earth~\cite{cowardin2024orbital, curzi2020large}.
Moreover, once a collision or breakup of satellites occurs, a lot of space debris is generated. 
Thus, space situational awareness (SSA) (or space domain awareness, SDA) aiming to detect, track, and understand the motion of Earth-orbiting RSOs, is becoming more important in managing the traffic of satellites.
For the advancement of SSA/SDA, much information such as orbit, attitude, and shape of RSOs enables improving the accuracy of state propagation of RSOs. 
The attitude of RSOs is especially a key state for improvement because non-gravitational forces such as solar radiation pressure and atmospheric drag depend on the object's attitude. 
In this context, this study focuses on the attitude estimation of an RSO via light curves.

Light curves are time variation of an object's brightness.
They are a function of the orbit, attitude, shape, and surface properties of the space object.
In other words, these characteristics can be inversely deduced from light curves.
This process is known as light curve inversion and is a promising and cost-effective method to estimate the RSO's attitude.
For example, it is difficult to take images of objects in geosynchronous orbit using ground-based equipment, but light curves can be observed.
Light curve inversion has been studied for a long time in the field of astronomy to determine rotation periods, rotation axis, and shape of asteroids~\cite{kaasalainen2001optimization, muinonen2020asteroid, muller2017hayabusa}.
Although asteroids are usually assumed to have a convex shape and diffusive reflection, the RSOs have non-convex shape and specular properties including anisotropic reflection, which significantly makes the light curve inversion of the RSOs challenging.
In previous studies, various attitude estimation methods using light curves have been proposed. 
For example, Wetterer et al.~\cite{wetterer2009attitude} show an attitude estimation method via light curves using an unscented Kalman filter (UKF). 
Cabrera et al.~\cite{cabrera2023adaptive} propose an adaptive Gaussian unscented Kalman filter, which enables capturing high non-linearity of the attitude estimation from light curves.
Du et al.~\cite{du2018attitude} present an attitude estimation method for geostationary satellites based on unscented particle filters.
Burton et al.~\cite{burton2023fast} offer an attitude estimation method that is robust against initial estimated values using a swarm optimizer and a quasi-Newton method.
Piergentili et al.~\cite{7819454} describe the attitude determination method using the 3-D virtual reality model and the genetic algorithm. 
Matsushita et al.~\cite{matsu} achieve an accurate attitude estimation using the sudden change of light curves, called glint.
However, most of the proposed methods assume that the target object's parameters such as shape and surface parameters are known completely, though they are usually unknown for space debris generated by a collision or breakup.
When the shape and surface parameters of the object are unknown, the 3-D virtual reality model and the light curve model for the observation step of the Kalman filters cannot be generated because they are parametrically formulated with the object’s orbit, attitude, shape, and surface parameters. 

This study tackles this problem by combining the Gaussian process regression (GPR) and the UKF, which is called GPUKF.  
The GPR is one of the machine learning methods that allows non-parametric regression~\cite{books/lib/RasmussenW06}. The UKF is one of the Bayesian filters and is used for filtering the noise and estimating the state of non-linear systems from observed values~\cite{wan2000unscented}. A conventional UKF consists of a prediction step and an observation step in the estimation sequence, which require parametric models. On the other hand, the GPUKF can represent the models non-parametrically by using the GPR. This study uses the GPR to describe the light curve model in the observation update steps of the estimation sequence, which require no information of shape and surface parameters. Thus, the object attitude can be estimated from the light curves with GPUKF even if the surface parameters are unknown. Furthermore, thanks to the non-parametric model, this estimation method is robust against uncertainties in those parameters.

Although the GPR would enable the description of both the system model and the light curve model in the prediction step and the observation step, respectively, this study focuses on the observation step of the light curve model. 
In this study, it is assumed that the target object is on a geosynchronous orbit since it is difficult to estimate the attitude of geostationary objects by images on the ground observation equipment.
It is also assumed that the time of period of the observation is 10 minutes. 
Hence, for simplicity, the relative position of the object--observer--Sun is fixed in numerical simulations.
In addition, a box-wing object is used for both the training and testing.

The organization of this paper proceeds as follows.
First, the necessary equations for the numerical simulation are reviewed, such as bidirectional reflectance distribution function (BRDF), light curve model, attitude representation, GPR, and UKF in Section~\ref{sec:pre}.
Second, the application of GPR with the UKF to attitude estimation is described in Section~3. 
Then, the configuration of the training data and testing data are described.
Finally, Section~4 shows the attitude estimation results of GPUKF compared with ones of a conventional UKF. 
The comparison shows that the GPUKF enables robust attitude estimation against uncertainties in the surface properties, contributing to practical scenarios in SSA/SDA where object parameters are unknown.

\section{Preliminaries}\label{sec:pre}

\subsection{Bidirectional Reflectance Distribution Function (BRDF)} 
\begin{figure}[tb] 
  \begin{minipage}[b]{0.48\columnwidth}
    \centering
  \includegraphics[width=\columnwidth]{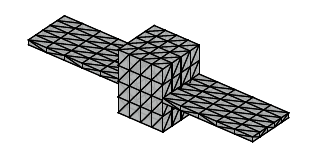}
  \subcaption{Facet model of target object.}
  \label{fig:boxWing_grided}
  \end{minipage}
  \begin{minipage}[b]{0.48\columnwidth}
    \centering
   \includegraphics[width=\columnwidth]{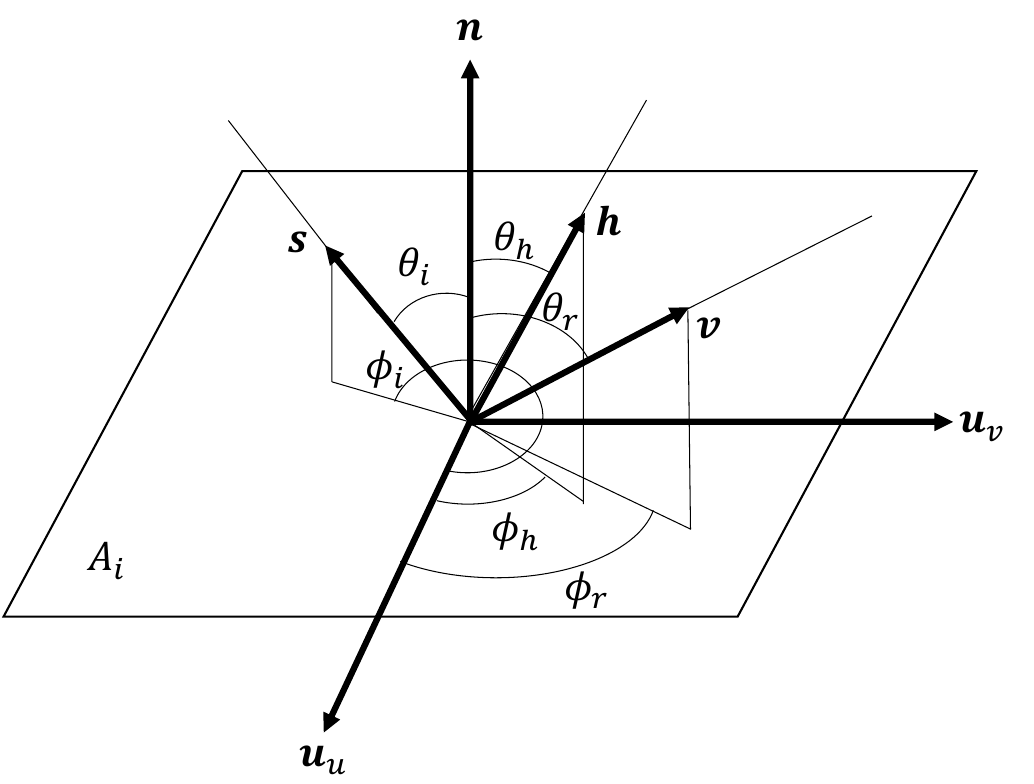}
     \subcaption{Light reflection on a facet.}
     \label{fig:facet_vector}
  \end{minipage}
  \caption{Facet model and geometric relation of light on each facet.}
\end{figure}

A BRDF $f_r$ is the function that describes the reflection property of an object~\cite{kwast2014introduction}.
A BRDF is represented as
\begin{equation}
    f_r(\bm{s},  \bm{v}) = \frac{L_r(\bm{v})}{L_i(\bm{s})} \label{eq:fr_1}
\end{equation}
where $L_r$ is the reflected radiance, $L_i$ is the irradiance, 
$\bm{s}$ is the unit vector of the light source direction, and $\bm{v}$ is the unit vector of the observer direction.
In this paper, the target object is split into many small facets as shown in Fig.~\ref{fig:boxWing_grided}. The light reflection is considered locally on each facet and their summation yields light curves in total.
The geometry of the light reflected on a facet is illustrated in Fig.~\ref{fig:facet_vector}, where the origin of the reference frame is the geometric center of the facet. 
This frame is a right-handed frame. The third axis $\bm{n}$ is the normal vector of the facet and the other axes $\bm{u}_u$ and $\bm{u}_v$ span the facet, which define the anisotropic direction.
The vectors $\bm{s}$ and $\bm{v}$ are locally expressed with respect to the facet frame with the sets of azimuth and elevation angles, $(\phi_i, \theta_i)$ and $(\phi_r, \theta_r)$, respectively. 
The ranges of azimuth and elevation angles are $(\phi_{i},\phi_{r})\in [0, 2\pi)$ and $(\theta_{i},\theta_{r}) \in [0, 2\pi]$. The bisector vector $\bm{h}$ between $\bm{s}$ and $\bm{v}$ is defined as
\begin{equation}
    \bm{h} = \frac{\bm{s} + \bm{v}}{\| \bm{s} + \bm{v} \|} = \begin{bmatrix}
     \sin\theta_h \cos\phi_h \\
     \sin\theta_h\sin\phi_h \\ 
     \cos\alpha
\end{bmatrix}
\end{equation}
where $\theta_h$ and $\phi_h$ are the elevation and the azimuth angle of $\bm{h}$.

The BRDF is also represented as
\begin{equation}
    f_r =  c_d +  c_s \label{eq:fr_2}
\end{equation}
In this study, the Ashikhmin-Shirley model~\cite{ashikhmin2000anisotropic} is used for the reflection model. The Ashikhmin--Shirley model is an anisotropoic model and can be described as
\begin{align}
    c_d &= \frac{28\rho}{23\pi} (1 - F_0) \left[ 1 - \left( 1 - \frac{\bm{n}^T\bm{s}}{2} \right)^5 \right] \left[ 1 - \left( 1 - \frac{\bm{n}^T\bm{v}}{1} \right)^5 \right] \label{eq:c_d_AS} \\
    c_s &= \frac{\sqrt{(n_u+1)(n_v+1)}}{8 \pi} \frac{F}{\bm{v}^T \bm{h} \max\{\bm{n}^T\bm{s}, \bm{n}^T\bm{v}\}} (\cos\alpha)^{\gamma}
\end{align}
where $\rho$ is the factor that specifies the diffuse reflectance of the substrate under the specular coating and 
$n_u$ and $n_v$ are the factors that define the shape of specular lobe along $\bm{u}_{u}$ and $\bm{u}_{v}$, respectively.
The exponent $\gamma$ and the Fresnel reflectance $F$ are defined as
\begin{align}
    \gamma &= n_u\cos^2\phi_h + n_v \sin^2\phi_h 
    = \frac{n_{u}(\bm{h}^{T}\bm{u}_{u})^{2} + n_{v}(\bm{h}^{T}\bm{u}_{v})^{2}}{1-(\bm{h}^{T}\bm{n})^{2}} \\
    F &= F_0 + \left( 1 - F_0 \right) \left( 1 - \bm{v}^T \bm{h} \right)^5 \label{eq:Fresnel_reflectance}
\end{align}
where $F_0$ is the material's reflectance for the normal incidence.
The range and unit of measurement of the parameters are shown in Table~\ref{tab:BRDF_para}.


\begin{table}[tbp]
    \centering
    \caption{Parameters for BRDF.}
    \begin{tabular}{lcc}
        \hline  \hline
        Parameter & Range & Unit \\
        \hline
        Radiance, $L_r$ & $\in \mathbb{R}$ & $\si{W/m^2 str}$ \\
        Irradiance, $L_i$ & $\in \mathbb{R}$ & $\si{W/m^2}$ \\
        factor of anisotropy, $n_u, n_v$ & $\in \mathbb{R}$ & - \\
        Fresnel reflectance, $F$ & $\in [0, 1]$ & - \\
        Material's reflectance, $F_0$ & $\in [0, 1]$ & - \\
        \hline  \hline
    \end{tabular}
    \label{tab:BRDF_para}
\end{table}

\subsection{Light Curves} 
Light curves are a series of the space objects' brightness observed by a telescope and their intensity depends on incoming light and reflected light.
The relative magnitude of the brightness of a facet observed on the ground is described as follows.
\begin{equation}
    m_\mathrm{app} = m_\mathrm{sun} - 2.5 \log_{10} \frac{f_\mathrm{obs}}{r_\mathrm{obs}^2} \label{eq:mApp}
\end{equation}
where $m_\mathrm{sun} = -26.7$ is the apparent magnitude of the brightness of the Sun and $r_\mathrm{obs}$ is the distance between the facet and the observer.
It is assumed that $r_\mathrm{obs}$ is the same for all the facets because the distance is much larger than the size of the object.
The magnitude term $f_\mathrm{obs}$ is written as
\begin{equation}
    f_\mathrm{obs} = f_{r} f_i 
    A (\bm{n}^T\bm{v}) \label{eq:f_obs}
\end{equation}
where $A$ is the area of the facet and $f_i$ 
is the projection of the incident sunlight along the normal unit vector of the facet, which is equal to
\begin{equation}
    f_i = 
    \begin{cases}
    0 & (\bm{n}^T \bm{s} \leq 0)\\
    \bm{n}^T \bm{s} & (\bm{n}^T \bm{s} > 0)
    \end{cases} \label{eq:fi}
\end{equation}

\subsection{Attitude Kinematics and Dynamics of a Space Object} 
\begin{figure}[tbp]
    \centering
    \includegraphics[width = \columnwidth]{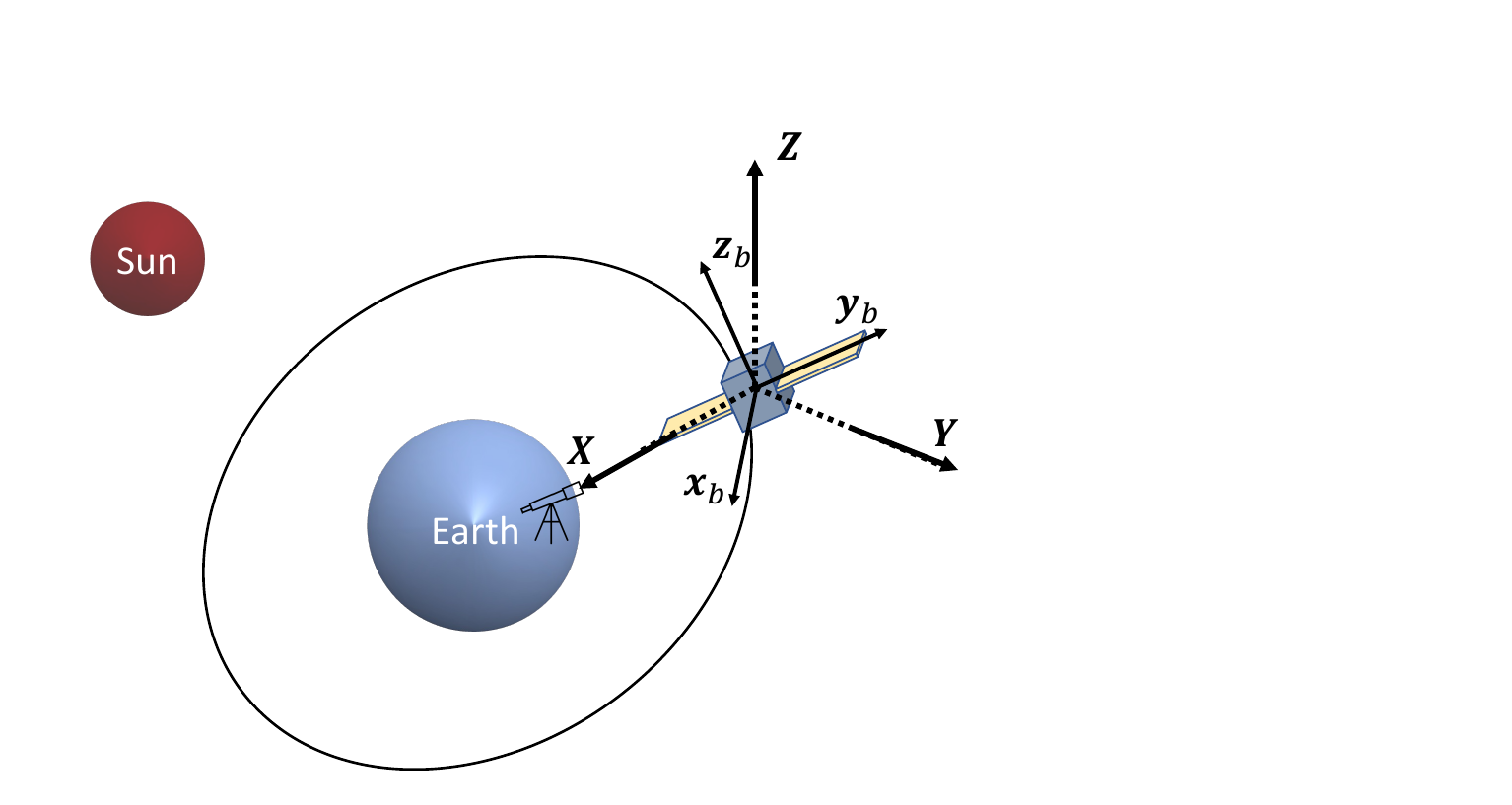}
    \caption{XYZ coordinate system and body-fixed frame.}
    \label{fig:coSys_frame}
\end{figure}

The coordinate system and the frame used in this study are shown in Fig.~\ref{fig:coSys_frame}.
The XYZ coordinate system
is a pseudo-inertial coordinate system and is used to represent the position of the space object orbiting Earth, the observer, and the Sun.
Since this study considers a fixed object-observer-Sun relative geometry for simplicity,
the pseudo-inertial frame is defined so that its origin is at the center of mass of the space object.
The $X$ axis points in the direction of the observer, $Z$ axis is along with the North pole of the Earth, and the $Y$ azis is defined to consist of the right-handed frame.
The body-fixed frame in Fig.~\ref{fig:coSys_frame} is used to represent the attitude of the space object.
It originates at the center of mass of the space object and each axis $\bm{x}_b$, $\bm{y}_b$, and $\bm{z}_b$ points towards the direction of the principal axes of inertia.

This paper uses quaternions to represent an attitude state with four parameters that include unit rotation vector (3 degrees of freedom) and rotation angle (1 degree of freedom) from Euler's theorem for rigid body rotation.
A quaternion is defined as
\begin{align}
    \bm{q} 
    &= [q_1, q_2, q_3, q_4]^T = [\bar{\bm{q}}^T, q_4]^T \nonumber \\
    &= \begin{bmatrix}
        \bm{e} \sin\frac{\Phi}{2} \\
        \cos\frac{\Phi}{2}
    \end{bmatrix} \label{eq:quaternion_def}
\end{align}
where $\bm{e}$ is a unit rotation vector, $\Phi$ is a rotation angle, and $\bar{\bm{q}}$ and $q_4$ are called vector part and scalar part of the quaternion, respectively.
The advantage of using quaternions for the attitude representation is that quaternions have no singularities.
However, it has the constraint that its norm is one as shown in Eq.~\eqref{eq:quaternion_constraint}.
\begin{align}
    \bm{q}^T \bm{q} = 1 \label{eq:quaternion_constraint}
\end{align}
This constraint causes that four arithmetic operations cannot be applied to the attitude calculation using quaternions.
Thus, the quaternion product is used to represent the change of attitude.
The quaternion product is defined as
\begin{align}
    \bm{q} \otimes \bm{q}' = \begin{bmatrix}
        q_4 \bar{\bm{q}}' + q'_4 \bar{\bm{q}} - \bar{\bm{q}} \times \bar{\bm{q}}' \\
        q_{4} q'_4 - \bar{\bm{q}}^T \bar{\bm{q}}'
    \end{bmatrix} \label{eq:quaternion_product}
\end{align}
Eq.~\eqref{eq:quaternion_product} means rotating the attitude represented by a quaternion $\bm{q}'$ by a quaternion $\bm{q}$.
The kinematics of the quaternion $\bm{q}$ is wrriten as
\begin{align}
    \frac{\mathrm{d}}{\mathrm{d}t}\left[\begin{array}{c}
    q_{1}\\
    q_{2}\\
    q_{3}\\
    q_{4}
    \end{array}\right] &= \frac{1}{2} \tilde{\bm{\omega}} \otimes \bm{q} \label{eq:quaternion_omega}
\end{align}
where $\bm{\omega}$ is an angular velocity and $\tilde{\bm{\omega}}=[\bm{\omega}^{T}, 0]^{T}$.

This paper uses the generalized Rodrigues parameters for a weighted average of the sigma points in the UKF because the generalized Rodrigues parameters have no constraint unlike the quaternions.
The generalized rodrigues parameters are defined as
\begin{align}
\bm{p} = f\frac{\bar{\bm{q}}}{a+q_{4}}
\end{align}
where $a$ is a parameter from 0 to 1 and $f$ is a scale factor.

\begin{figure}[tbp]
    \centering
    \includegraphics[width = 0.3\columnwidth]{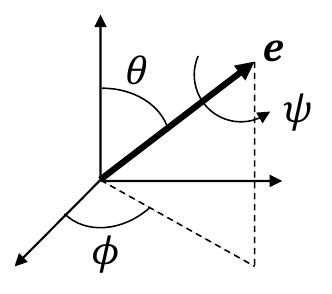}
    \caption{Rotation axis unit vector.}
    \label{fig:azi_polar}
\end{figure}

Euler angles are an alternative and widely used way of representing a definition of representing the orientation of an object. 
This paper uses them to show the results of numerical simulations.
Euler angles represent the attitude by conducting three consecutive rotations around each coordinate axis.
Thus, the attitude representation can be understood intuitively.
The rotation matrices around each axis of a frame are written as $R_z(\phi)$, $R_y(\theta)$, and $R_x(\psi)$ 
where $\phi$ (yaw), $\theta$ (pitch), and $\psi$ (roll) are rotation angles of each axis, respectively.
The rotation matrix of the Euler angle (3-2-1 sequence) from the inertial frame ($i$-frame) to the body-fixed frame ($b$-frame) is described as
\footnotesize
\begin{align}
R_{b/i} & =R_{x}(\psi)R_{y}(\theta)R_{z}(\phi) \nonumber \\
&= \left[\begin{array}{ccc}
\cos\theta\cos\phi & \cos\theta\sin\phi & -\sin\theta\\
\sin\theta\sin\psi\cos\phi-\cos\psi\sin\phi & \sin\theta\sin\psi\sin\phi+\cos\psi\cos\phi & \cos\theta\sin\psi\\
\sin\theta\cos\psi\cos\phi+\sin\psi\sin\phi & \sin\theta\cos\psi\sin\phi-\sin\psi\cos\phi & \cos\theta\cos\psi
\end{array}\right] \label{eq:R_b_i}
\end{align}
\normalsize
Furthermore, an attitude can be represented by the unit vector of the rotation axis $\bm{e}$ and the rotation angle $\Phi$ shown in Fig.~\ref{fig:azi_polar}.
The relational expression between the rotation vector and the quaternions is written as
\begin{equation}
    \bm{q} = \left[ \begin{array}{c}
    \sin\theta \cos\phi \sin\frac{\psi}{2}\\ \sin\theta \sin\phi \sin\frac{\psi}{2} \\ \cos\theta \sin\frac{\psi}{2} \\
    \cos\frac{\psi}{2}
    \end{array} \right]
\end{equation}

A space object is assumed to be a rigid body, and Euler's equation is used to represent the object's attitude motion.
It follows the conservation of angular momentum as
\begin{align}
    \frac{d\bm{h}}{dt} &= \frac{d^* \bm{h}}{dt} + \bm{\omega} \times \bm{h}  \nonumber \\ 
    & = J \frac{d^* \bm{\omega}}{dt} + \bm{\omega} \times J \bm{\omega} = \bm{\tau} \label{eq:euler_equation}
\end{align}
where $\bm{h}$ is the angular momentum vector, $J$ is the moment of inertia, $\bm{\tau}$ is the external torque, 
$\frac{d\bm{h}}{dt}$ and $\frac{d^*}{dt}$ is the time differential with respect to the body-fixed inertial frame and the body-fixed frame respectively.

\subsection{Gaussian Process Regression (GPR)}
Gaussian process regression (GPR) is one of the machine learning methods that non-parametrically models non-linear systems~\cite{books/lib/RasmussenW06}. 
The regression model of the function $y_\mathrm{gp} = \mathrm{GP}(\bm{x}_\mathrm{gp})$ has the scalar output $y_\mathrm{gp}$ from the input vector $\bm{x}_\mathrm{gp}$ with dimension $m$. 
A training data set is represented as
\begin{align}
    \mathcal{D} = \{ (\bm{x}_{\mathrm{gp}, 1}, y_{\mathrm{gp}, 1}), (\bm{x}_{\mathrm{gp}, 2}, y_{\mathrm{gp}, 2}), \ldots, (\bm{x}_{\mathrm{gp}, N}, y_{\mathrm{gp}, N}) \} = \langle X_\mathrm{gp}, \bm{y}_\mathrm{gp} \rangle \label{eq:D}
\end{align}
where $\bm{x}_{\mathrm{gp}, k}~(k=1,\dots,N)$ is the input vector and $y_{\mathrm{gp}, k}$ is the scalar output of the corresponding input $\bm{x}_{\mathrm{gp}, k}$.
The output vector $\bm{y}_\mathrm{gp} = [y_{\mathrm{gp}, 1}, y_{\mathrm{gp}, 2}, \ldots, y_{\mathrm{gp}, N}]^T$ is the multivariate Gaussian distribution normalized to have a zero mean and its covariance matrix $K$ as
\begin{equation}
    \bm{y}_\mathrm{gp} \sim \mathcal{N}(0, K) 
\end{equation}
where $K$ is the $N \times N$ kernel matrix of the input data. 
The element $K_{nn'}$ is defined as
\begin{equation}
K_{nn'} = k(\bm{x}_{\mathrm{gp}, n}, \bm{x}_{\mathrm{gp}, n'}) + \epsilon^2 \label{eq:K_element}    
\end{equation}
where $k(\cdot)$ is the kernel function and $\epsilon^2$ is the variance of the observation noise.
In this study, the Gaussian kernel represented by Eq.~\eqref{eq:gaussian_kernel} is used.
\begin{equation}
    k(\bm{x}_{\mathrm{gp}, n},\bm{x}_{\mathrm{gp}, n'}) = \theta_1 \exp \left( - \frac{\|\bm{x}_{\mathrm{gp}, n} - \bm{x}_{\mathrm{gp}, n'}\|^2}{\theta_2} \right) \label{eq:gaussian_kernel}
\end{equation}
where $\theta_{1}$, $\theta_{2}$ are hyperparameters of the GPR, and $\bm{x}_{\mathrm{gp}, n}$, $\bm{x}_{\mathrm{gp},n'}$ are two kinds of inputs vector for the kernel function.

Given a new input vector $\bm{x}^*_\mathrm{gp}$, the predictive distribution of $y^*_\mathrm{gp}$ is represented as
\begin{equation}
     p(y^*_\mathrm{gp} | \bm{x}^*_\mathrm{gp}, \mathcal{D}) = \mathcal{N} (\bm{k}^T_* K^{-1} \bm{y}_\mathrm{gp}, k_{**} - \bm{k}^T_* K^{-1} \bm{k}_*) \label{eq:gpr_predictive_distribution}
\end{equation}
That is, the output $y^*_\mathrm{gp}$ corresponding to the input $\bm{x}^*_\mathrm{gp}$ can be obtained from the Gaussian process trained by data set $\mathcal{D}$.
The output $y^*_\mathrm{gp}$ regressed by GP is denoted as
\begin{align}
    y^*_\mathrm{gp} = \mathrm{GP}(\bm{x}^*_\mathrm{gp}, \mathcal{D})
\end{align}
where the mean and the variance of $y^*_\mathrm{gp}$ are respectively denoted as
\begin{align}
    \mathrm{GP}_\mu(\bm{x}^*_\mathrm{gp}, \mathcal{D}) &= \bm{k}_*^T K^{-1} \bm{y}_\mathrm{gp} \label{eq:GP_mu} \\
    \mathrm{GP}_\Sigma (\bm{x}^*_\mathrm{gp}, \mathcal{D}) &= k_{**} - \bm{k}_*^T K^{-1} \bm{k}_* \label{eq:GP_Sigma}
\end{align}
where
\begin{align}
    \bm{k}_* &= [k(\bm{x}^*_\mathrm{gp}, \bm{x}_{\mathrm{gp}, 1}), k(\bm{x}^*_\mathrm{gp}, \bm{x}_{\mathrm{gp}, 2}), \ldots, k(\bm{x}^*_\mathrm{gp}, \bm{x}_{\mathrm{gp}, N})]^T \label{eq:k_*} \\
    k_{**} &= k(\bm{x}^*_\mathrm{gp}, \bm{x}^*_\mathrm{gp}) \label{eq:k_**}
\end{align}
The algorithm of GPR is summarized in Algorithm~\ref{alg:gpr}.

\begin{figure}[tbp]
\begin{algorithm}[H]
    \caption{The GPR algorithm.}
    \label{alg:gpr}
    \begin{algorithmic}[1]
        \STATE $[mu, var] = gpr(xtest, xtrain, ytrain, kernel)$
        \STATE $N = length(ytrain)$
        \FOR{$n = 1, \ldots, N$}
            \FOR{$n' = 1, \ldots, N$}
                \STATE $K[n, n'] = kernel(xtrain[n], xtrain[n'])$
            \ENDFOR
        \ENDFOR
        \STATE $yy = K^{-1} * ytrain$
        \FOR{$m = 1 , \ldots, M$}
            \FOR{$n = 1, \ldots, N$}
                \STATE $k[n] = kernel(xtrain[n], xtest[m])$
            \ENDFOR
            \STATE $s = kernel(xtest[m], xtest[m])$
            \STATE $mu[m] = k^T * yy$
            \STATE $var[m] = s - k^T * K ^{-1} * k$
        \ENDFOR
    \end{algorithmic}
\end{algorithm}
\end{figure}


\subsection{Unscented Kalman Filter (UKF)}
A discrete-time nonlinear system model is given by 
\begin{align}
    \bm{x}_{k+1}&=\bm{f}(k, \bm{x}_{k}, \bm{v}_{k},\bm{u}_{k}) \in \mathbb{R}^{m} \label{eq:ukf_prediction_model} \\ 
\bm{y}_{k}&=\bm{h}(k, \bm{x}_{k}, \bm{\epsilon}_{k},\bm{u}_{k}) \in \mathbb{R}^{l} \label{eq:ukf_observation_model}
\end{align}
where $\bm{x}_k$ is the $m \times 1$ state vector, $\bm{u}_k$ is the input vector, $\bm{y}_k$ is the $ l \times 1$ observation vector at discrete-time $t_k$, and
$\bm{v}_k$ and $\bm{\epsilon}_k$ are the process noise and the observation noise, respectively.
The UKF approximates a Gaussian distribution using sigma points that are obtained as
\begin{align}
\bm{\chi}_{k-1,0} &= \hat{\bm{x}}_{k-1} \label{eq:sigmapoints_x_1} \\ 
\bm{\chi}_{k-1,i} &= \bm{\chi}_{k-1,0} + \bm{\sigma}_{k-1,i}\hspace{10pt}(i=1,2,\dots,2m) \label{eq:sigmapoints_x_2} \\
\bm{\sigma}_{k-1} &= [\sqrt{m + \lambda} \sqrt{P_{k-1}}, -\sqrt{m + \lambda} \sqrt{P_{k-1}}] \in \mathbb{R}^{m \times 2m} \label{eq:sigmapoints_x_3}
\end{align}
where $\bm{\chi}_{k-1, i}~(i=0,1,\dots,2m)$ are the sigma points, $\hat{\bm{x}}_{k-1}$ is the state estimated at $t_{k-1}$, $P_k = \sqrt{P_k} \sqrt{P_k}^T$ is the $m \times m$ covariance matrix of the $\bm{x}_k$. 
The scaling parameter $\lambda$ is defined as 
\begin{align}
    \lambda = \alpha^{2}\left(m+\kappa \right) - m
\end{align}
where $\kappa = 3 - m$ and $10^{-4} < \alpha < 1$ are tuning parameters.
The substitution of sigma points $\bm{\chi}_{k-1}$ into Eq.~\eqref{eq:ukf_prediction_model} yields predicted sigma points $\bm{\chi}^-_{k}$ as
\begin{align}
    \bm{\chi}^{-}_{k} = \bm{f}(k-1,\bm{\chi}_{k-1},\bm{u}_{k-1})
\end{align}
Taking a weighted average of the predicted sigma points $\bm{\chi}^-_{k}$, a priori predicted state value and covariance matrix are obtained as follows.
\begin{align}
    \hat{\bm{x}}^{-}_{k} &= \sum_{i=0}^{2m} w_{i,{\rm mean}} \bm{\chi}_{k,i}^{-} \label{eq:priori_predicted_state}\\
    P^{-}_{k} &= \sum_{i=0}^{2m}w_{i,{\rm cov}} \left(\bm{\chi}_{k,i}^{-} - \hat{\bm{x}}_{k}^{-}\right)\left(\bm{\chi}_{k,i}^{-} - \hat{\bm{x}}_{k}^{-}\right)^{T} \label{eq:priori_predicted_cov}
\end{align}
The weights $w_{i, {\rm mean}}$ and $w_{i, {\rm cov}}$ are given by
\begin{align}
    w_{0,{\rm mean}} &= \frac{\lambda}{m + \lambda} \\
    w_{0,{\rm cov}} &= \frac{\lambda}{m + \lambda} + \left(1-\alpha^{2}+\beta\right) \\
    w_{i,{\rm mean}} &= w_{i,{\rm cov}} = \frac{1}{2\left(m + \lambda\right)}, \quad (i=1,2,\dots, 2m)
\end{align}
where the parameter $\beta$ is the tuning parameter about the distribution of the estimated state. 
When the distribution is assumed to be the Gaussian distribution, $\beta = 2$ is appropriate~\cite{10.5555/1121596}.
The Sigma points $\hat{\bm{\chi}}^-_k$ in Eq.~\eqref{eq:priori_predicted_state} are computed again and used for the calculation of a priori predicted observation $\hat{\bm{y}}^-_k$ as
\begin{align}
    \bm{\mathcal{Y}}^-_{k,i} &= \bm{h}(\bm{\chi}_{k,i}^{-}), \hspace{10pt} (i=0,1,\dots, 2m) \label{eq:sigmapoints_y} \\
    \hat{\bm{y}}^{-}_k & = \sum_{i=0}^{2m}w_{i, {\rm mean}}\bm{\mathcal{Y}}^-_{k,i} \label{eq:priori_predicted_observed_value}
\end{align}
The Kalman gain is given by
\begin{align}
    G_{k} = P_{k}^{xy} \left(P_{k}^{yy}\right)^{-1} \label{eq:kalman_gain}
\end{align}
where covariance matrices $P_k^{xy}$ and $P_k^{yy}$ are defined as
\begin{align}
P_{k}^{yy} &= \sum_{i=0}^{2m}w_{i,{\rm cov}}\left(\bm{\mathcal{Y}}^-_{k,i} - \hat{\bm{y}}^{-}\right)\left(\bm{\mathcal{Y}}^-_{k,i}-\hat{\bm{y}}^{-}\right)^{T} \in \mathbb{R}^{l \times l}\\
P^{xy}_{k} &= \sum_{i=0}^{2m}w_{i,{\rm cov}}\left(\bm{\chi}_{k}^{i,-}-\hat{\bm{x}}^{-}\right)\left(\bm{\mathcal{Y}}^-_{k,i}-\hat{\bm{y}}^{-}\right)^{T} \in \mathbb{R}^{m \times l}
\end{align}
Consequently, posterior estimated state value $\hat{\bm{x}}^+_k$ and covariance matrix $P_k^+$ are obtained as follows.
\begin{align}
\hat{\bm{x}}^{+}_k &= \hat{\bm{x}}_{k}^{-}+G_{k}\left(\tilde{\bm{y}}_{k}-\hat{\bm{y}}_{k}^{-}\right) \label{eq:posterior_state_value} \\
P^{+}_{k}&=P^{-}_{k}-G_{k}P_{k}^{yy}G_{k}^{T} \label{eq:posterior_cov}
\end{align}
where $\tilde{\bm{y}}$ is the vector of observed values at time $t_k$.

\section{Applying GPUKF to Attitude Estimation via Light Curves}
Bayesian filters, such as the extended Kalman filter and UKF, are often used for the estimation of the attitude via light curves. 
Such filters need a parametric state space model of a target object to construct a prediction model and an observation model, as shown in Eqs.~\eqref{eq:ukf_prediction_model} and~\eqref{eq:ukf_observation_model} respectively. 
In the case of attitude estimation from photometric data of a space object with unknown surface parameters, the observation model 
cannot be fully described.
To tackle this problem, this paper proposes an observation model for light curve inversion using GPR, enabling a non-parametric observation model. Furthermore, since the GPR is configured by training data with a variety of surface parameters, the estimation method using GPR is robust against the uncertainty in surface parameters.
The following subsections describe how the GP model is combined with the UKF for attitude estimation via light curves.

\subsection{Prediction Model and Observation Model}
\begin{figure}
    \centering
    \includegraphics[width=0.8\linewidth]{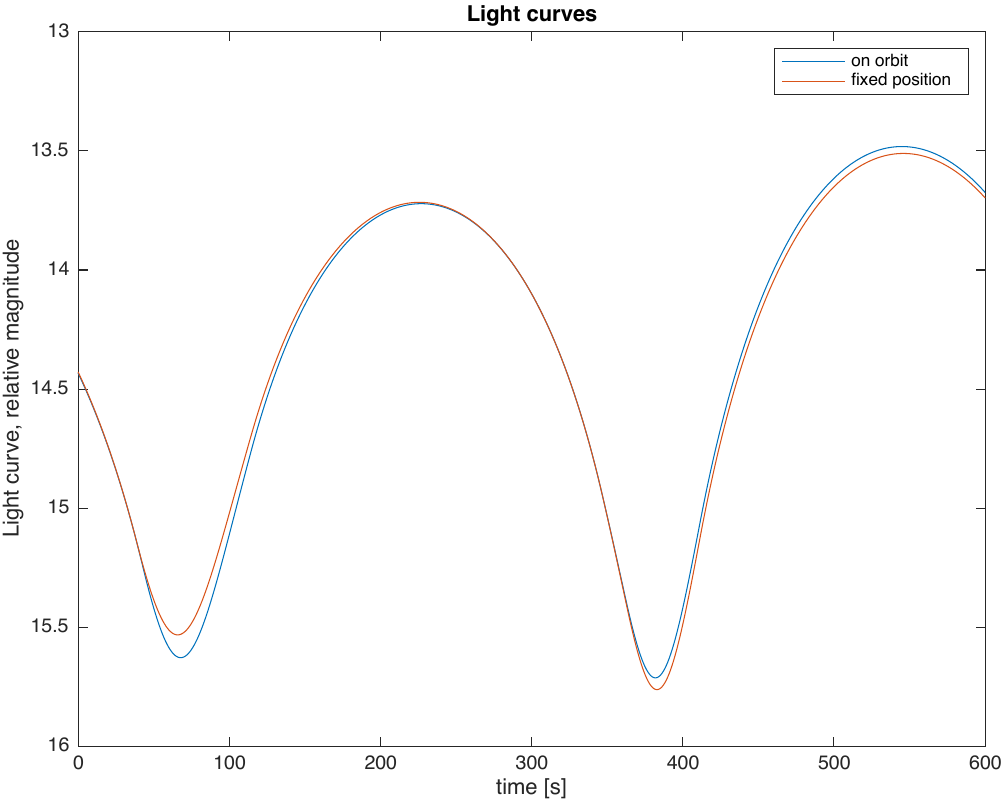}
    \caption{Comparison of the light curves with orbital propagation and with no orbital propagation.}
    \label{fig:cmp_lc_orbital_fixed}
\end{figure}

The prediction model estimates the attitude at the next time step from the current state vector. 
The observation model calculates the light curve from the current state vector.
The two models can be written as
\begin{align}
    \bm{x}_{\mathrm{prd}, k+1} &= \bm{f}(\bm{x}_{\mathrm{prd}, k}) \label{eq:prediction_model} \\
    m_{\mathrm{app}, k} &= \mathrm{GP}(\bm{x}_{\mathrm{obs}, k}, \mathcal{D}_\mathrm{obs}) \label{eq:observation_model}
\end{align}
where $\bm{f}$ is the function of the dynamics of the object, $\bm{x}_\mathrm{prd}$ is the input vector for the prediction model,  $\bm{x}_\mathrm{obs}$ is the input vector for the observation model, and $m_{\mathrm{app}, k}$ is the synthetic light curve.
This study uses the Euler's equation and the quaternion kinematics in Eqs.~\eqref{eq:euler_equation} and \eqref{eq:quaternion_omega} as the prediction model. The object--observer--Sun position is assumed to be fixed.
This assumption is appropriate because the target object is assumed to be on a geosynchronous orbit and the observation is for several minutes.
In fact, the light curves generated with the fixed position are almost the same as those with the orbital propagation, as shown in Fig.~\ref{fig:cmp_lc_orbital_fixed}.
In addition, the initial angular velocity is assumed to be known in this paper. Extending the proposed method would enable including the angular velocity in the training data to infer the angular velocities as well. 
As a first step towards a more practical methodology, this study considers quaternions and mean light curves because of the limitation of the memory of the computer.
The training data set $\mathcal{D}_\mathrm{obs}$ is the pairs of input values $\bm{x}_{\mathrm{obs}, k}$ and output value $m_{\mathrm{app}, k}$.
The details of the synthetic training data are described in Subsection~\ref{subsec:TrainingData}.

\subsection{Training Data} \label{subsec:TrainingData}

\begin{figure}[tbp]
    \centering
    \includegraphics[width = 0.5\columnwidth]{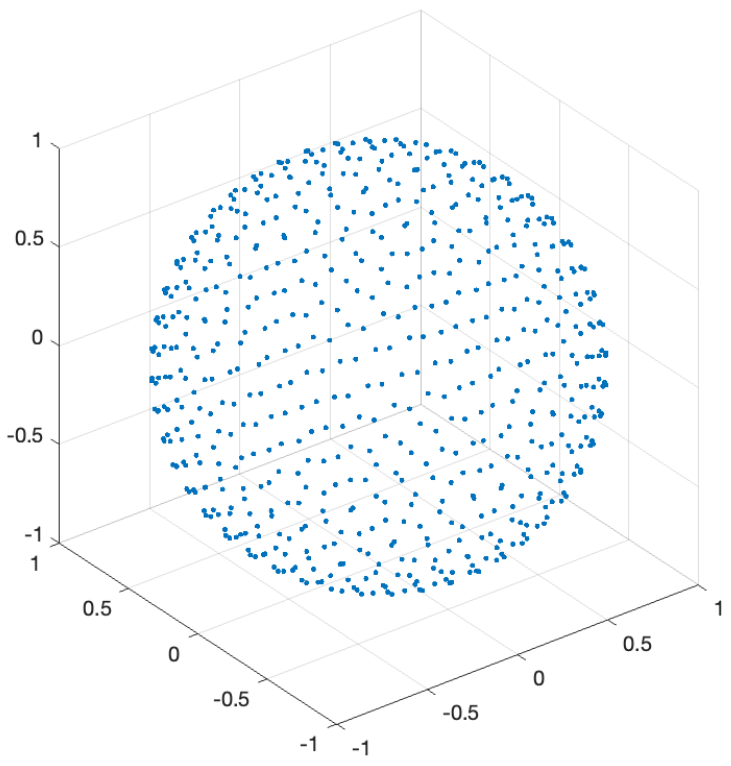}
    \caption{Unit vectors of rotational axes for entirely-covered state space of attitude.}
    \label{fig:icosphere_verticiesOnly}
\end{figure}

\begin{figure}
    \centering
    \includegraphics[width=1.0\linewidth]{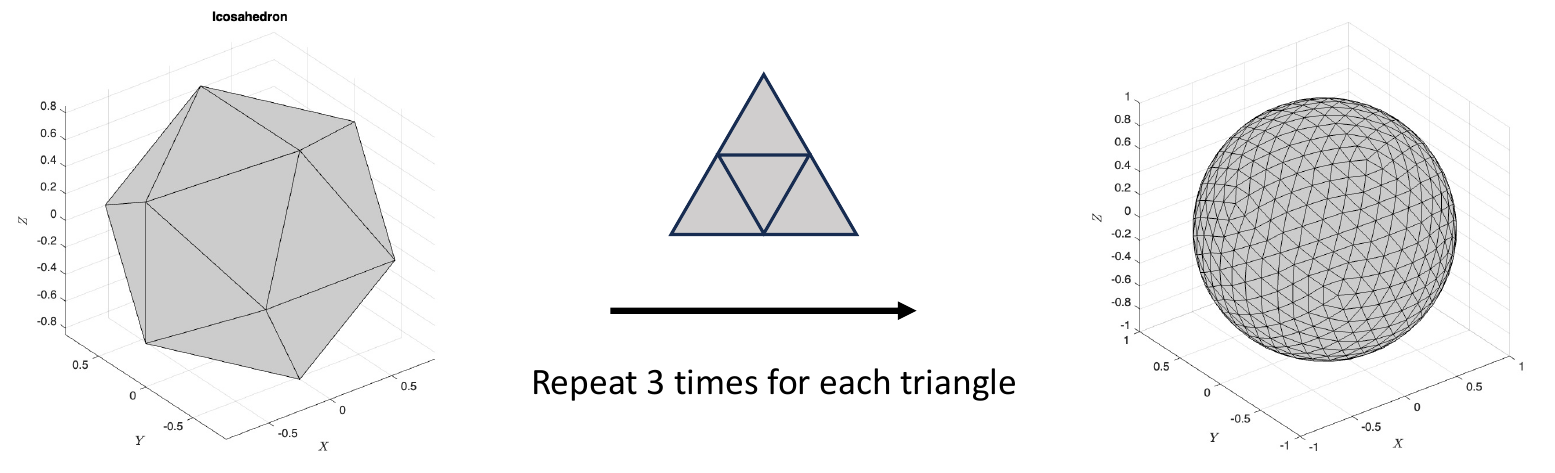}
    \caption{Generate polyhedron for rotation axes from icosahedron}
    \label{fig:generate_icosphere}
\end{figure}

The training data set for regression of the light curves is written as
\begin{equation}
    \mathcal{D}_\mathrm{obs} = \langle X_\mathrm{obs}, Z_\mathrm{obs} \rangle \label{eq:Do} 
\end{equation}
where $X_\mathrm{obs}$ is the set of inputs and $Z_\mathrm{obs}$ is the set of outputs.
They are defined as
\begin{align}
    X_\mathrm{obs} &= [\bm{x}_{\mathrm{obs}, 1}, \bm{x}_{\mathrm{obs}, 2}, \ldots, \bm{x}_{\mathrm{obs}, N}] \\
    Z &= [m_{\mathrm{app}, 1}, m_{\mathrm{app}, 2}, \ldots, m_{\mathrm{app}, N}]
\end{align}
The input vector $\bm{x}_{{\rm obs},k}$ in this paper is defined as
\begin{align}
    \bm{x}_{{\rm obs}, k} &= [\bm{q}_{k}^T, m_{\mathrm{app, past}, k} ]^T
\end{align}
where $\bm{q}_k$ is the quaternion of an object and $m_{\mathrm{app, past}, k}$ denotes the mean of the past light curves over $t_\mathrm{span, past}$ seconds, which are obtained by Eqs.~\eqref{eq:quaternion_omega}, \eqref{eq:euler_equation}, and \eqref{eq:mApp}.
The training data set must cover the entire attitude state space because accurate regrssion requires training data for many possible attitude angles.
Such training data can be obtained by varying the unit vector of the rotation axis in all directions as shown in Fig.~\ref{fig:icosphere_verticiesOnly} and varying the rotation angle $\Phi$ for each rotation axis.
These rotation axes for all directions are obtained from the regular polyhedron as shown in Fig.~\ref{fig:generate_icosphere}.
This regular polyhedron is based on an icosahedron. 
It is generated by repeating 3 times to split each triangle into 4 smaller triangles and normalizing the distance from the origin to each vertex.
Thus, the number of the rotation axes is 642.



It is noted that $m_{\mathrm{app, past}, k}$ is required in the input data to include information about surface parameters.
This is because multiple combinations of surface parameters and attitude can produce identical light curves, making the regression problem non-unique.
When the input vector consists of only the attitude quaternion, it captures the instantaneous information depending on the attitude and surface parameters.
On the other hand, using $m_{\mathrm{app, past}, k}$ in the input vector provides temporal variations in light curves that reflect both attitude changes and surface properties of the object.
Thus, the change in $m_{\mathrm{app}, k}$ due to different surface properties can be trained.
Furthermore, the mean of the past light curves can mitigate the influence of observation noise.
Therefore, the observation model in Eq.~\eqref{eq:observation_model} is expected to be robust against the uncertainty of the surface properties.
The initial angular velocity $\bm{\omega}_0$ is also set for propagation to obtain the time history of the attitude, and then the light curve corresponding to each attitude is calculated by Eq.~\eqref{eq:mApp}.
The final attitude and light curve in the time history are denoted as $\bm{q}_k$ and $m_{\mathrm{app}, k}$.


\subsection{GPUKF for Attitude Estimation via Light Curves}
\begin{figure}[tbp]
            \centering
            \includegraphics[width=0.9\linewidth]{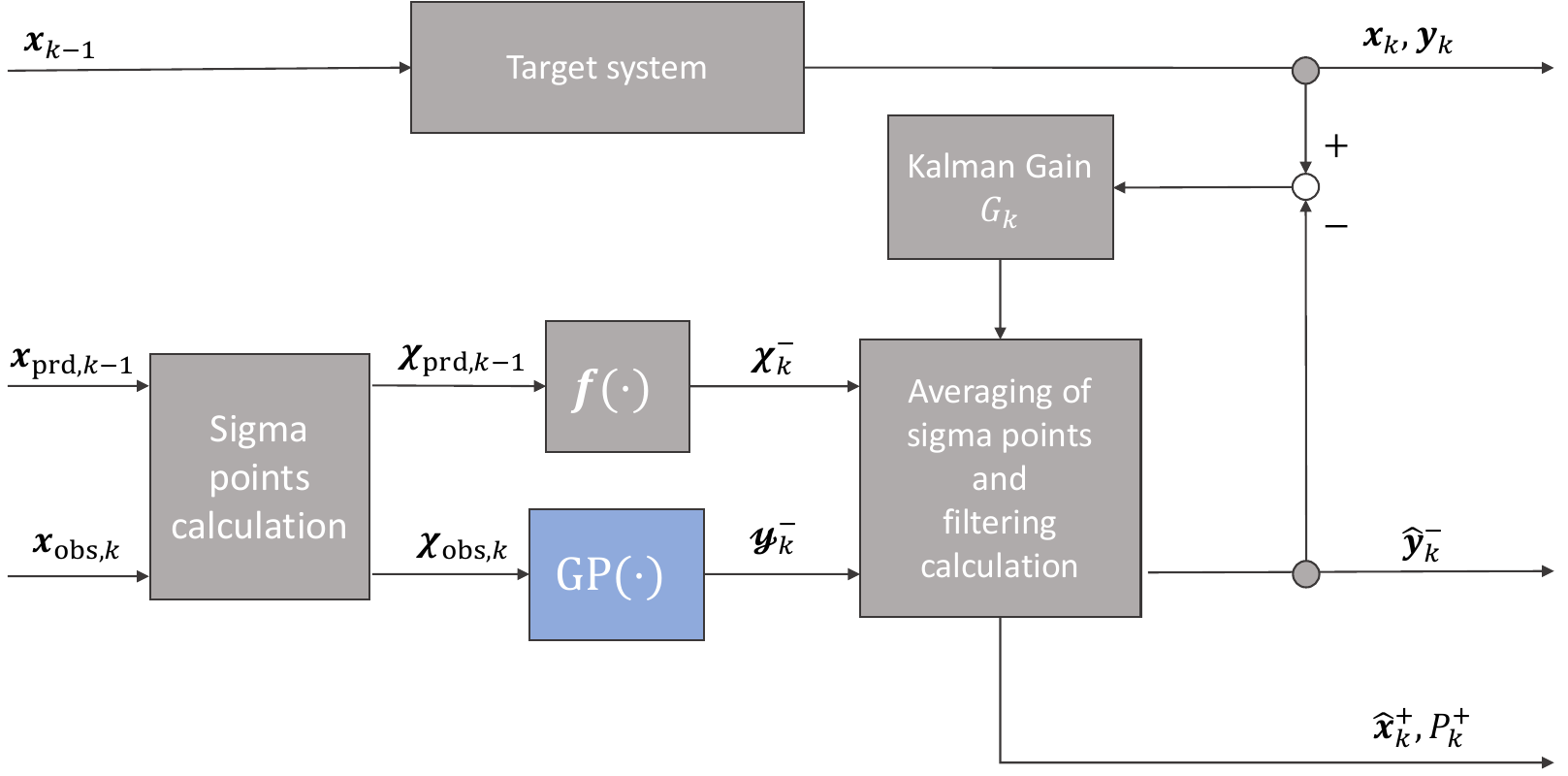}
            \caption{Algorithm of GPUKF.}\label{fig:GPUKF_algorithm}
\end{figure}

The algorithm of the GPUKF model for attitude estimation is shown in Fig.~\ref{fig:GPUKF_algorithm}.
First, the sigma points in Eqs.~\eqref{eq:sigmapoints_x_1}--\eqref{eq:sigmapoints_x_3} are calculated with the initial estimate, and they are substituted into the prediction model in Eq.~\eqref{eq:prediction_model}. 
Note that although the (global) attitude is represented with quaternions, the (local) attitude error is described with the generalized Rodrigues parameter (GRP) as proposed in~\cite{crassidis2003unscented}.  Thus, the sigma points are calculated for the GRP.
Then, the sigma points of the predicted observed value in Eq.~\eqref{eq:sigmapoints_y} are calculated by substituting the sigma points of the predicted state into the observation model of the GP in Eq.~\eqref{eq:observation_model}.
Finally, the Kalman gain in Eq.~\eqref{eq:kalman_gain} is calculated and the posterior estimated state value $\hat{\bm{x}}^+_k$ and the posterior covariance matrix $P^+_k$ are obtained by Eqs.~\eqref{eq:posterior_state_value} and \eqref{eq:posterior_cov}.
Additionally, $m_{\mathrm{app, past}, k}$, one of the input values for the GPUKF, is also updated by
\begin{equation}
    m_{\mathrm{app, past},k+1} = \frac{t_\mathrm{span, past} m_{\mathrm{app, past},k} + m_{\mathrm{app},k}}{t_\mathrm{span, past} + 1} \label{eq:m_appPast_updated}
\end{equation}
where $m_{\mathrm{app}, k}$ is the observed value at time $t_{k}$ and $t_\mathrm{span, past}$ is the time period for the averaging of the past light curves.

\section{Numerical Simulation}
\subsection{Parameters for Training and testing Data}
\begin{figure}
    \centering
    \includegraphics[width=0.7\linewidth]{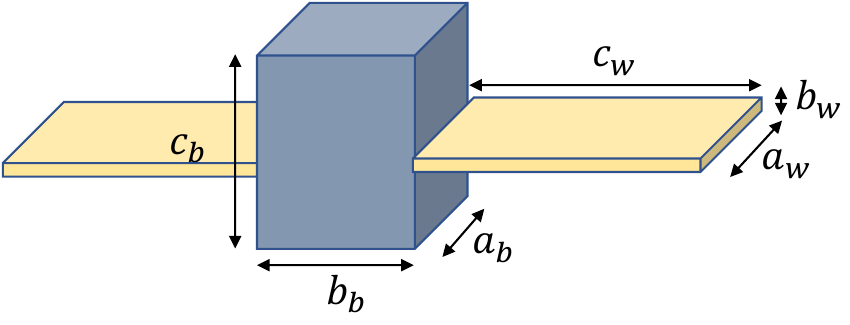}
    \caption{Box-Wing model.}
    \label{fig:boxWing}
\end{figure}

\begin{figure}[tbp]
    \centering
    \begin{minipage}[b]{0.45\linewidth}
        \centering
        \includegraphics[width=\linewidth]{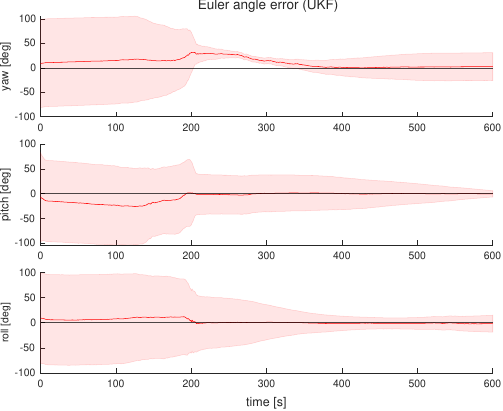}
        \subcaption{$\rho_\mathrm{ukf} = \rho_\mathrm{true} = 0.5$.}
        \label{fig:ukf_rho05_eulerAngleError}
    \end{minipage}
    \hfill
    \begin{minipage}[b]{0.45\linewidth}
        \centering
        \includegraphics[width=\linewidth]{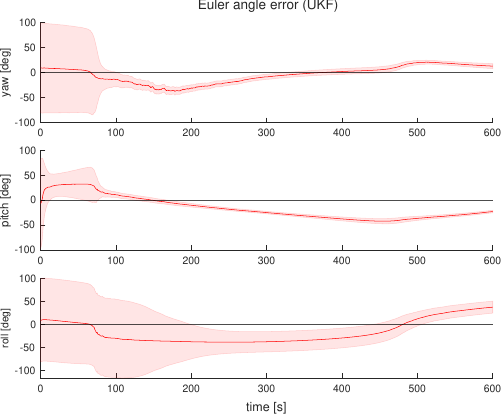}
        \subcaption{$\rho_\mathrm{ukf} = 0.5, \rho_\mathrm{true} = 0.7$.}
        \label{fig:ukf_rho_07_eulerAngleError}
    \end{minipage}
    \caption{Comparison between $\rho_\mathrm{true}$ and $\rho_\mathrm{ukf}$ using a conventional UKF.}
    \label{fig:ukf_doesNotWork_results}
\end{figure}

Numerical simulations are conducted with the box-wing satellite illustrated in Fig.~\ref{fig:boxWing} and its parameters are summarized in Table~\ref{tab:shape_para}. 
The relative position of the space object, observer, and Sun is written in Table~\ref{tab:obj-obs-sun_position_fixed}.
The phase angle, the angle between the Sun and the space object at the observer, is 29.8478~deg.
\begin{table}[tbp]
    \centering
    \caption{Target Object Parameters.}
    \begin{tabular}{lc}
        \hline  \hline
        Parameter & Value \\
        \hline
        Box side length, $(a_b, b_b, c_b)$ & $(1.0, 1.0, 1.3) ~\si{m}$ \\
        Wing side length, $(a_w, b_w, c_w)$ & $(1.0, 0.1, 2.0) ~\si{m}$ \\
        Moment of inertia , $(J_x, J_y, J_z)$ & $(55, 40, 65) ~\si{kgm^2}$ \\
        Number of facets & 472 \\
        \hline  \hline
    \end{tabular}
    \label{tab:shape_para}
\end{table}
\begin{table}[tbp]
    \centering
    \caption{Object--Observer--Sun relative position in the body-fixed inertial frame.}
    \footnotesize
    \begin{tabular}{lc}
        \hline \hline
        Parameter & Value \\
        \hline 
        Object position & $(0, 0, 0) ~\si{km}$ \\
        Observer position& $(36000, 0, 0)~\si{km}$ \\
        Sun position& $(1.2975, -0.3461, -0.6592) \times 10^8 ~\si{km}$ \\
        \hline \hline
    \end{tabular}
    \normalsize
    \label{tab:obj-obs-sun_position_fixed}
\end{table}
This study proposes an estimation method using GPUKF and investigates the robustness against uncertainties in surface properties.
Fig.~\ref{fig:ukf_doesNotWork_results} shows the attitude estimation error with a conventional UKF for different surface parameters. Note that the attitude error is described with Euler angles of 3-2-1 sequence. Fig.~\ref{fig:ukf_rho05_eulerAngleError} is the estimation error when the surface parameters in the UKF are the same as the true values, that is, $\rho_\mathrm{ukf}=\rho_\mathrm{true} = 0.5$, which successfully converges to zero. 
On the other hand, the estimation error in Fig.~\ref{fig:ukf_rho_07_eulerAngleError} is the result when one of the surface parameters $\rho = 0.5$ used in the UKF is different from the true one $\rho_\mathrm{true}=0.7$, which clearly deteriorates the estimation accuracy.
As it can be seen from these results, the uncertainty of surface parameters significantly affects the estimation error of UKF.
Thus, this study proposes an estimation method with GPUKF, where the training data set $\mathcal{D}_\mathrm{obs}$ for the observation model is generated for various surface parameters so that robustness against uncertainties in surface properties is achieved in the estimation of the attitude.


In the following simulations, the surface parameters of the Ashikhmin--Shirley model used for the training data are summarized in Table~\ref{tab:surface_para}, where the diffuse factor $\rho$ can assume various values. The initial attitudes are set to cover the state space entirely using the rotation axis unit vectors as shown in Fig.~\ref{fig:icosphere_verticiesOnly} and applying rotation angles $\Phi$ in increments of 60~deg from 0 to 300~deg.
Thus, the number of the initial attitude states for training is $642 \times 6 = 3852$.
The number of $\rho$ for the training dataset is 5 as shown in Table~\ref{tab:surface_para}.
Therefore, the number of the training data $N$ is $3852 \times 5 = 19260$.
The initial angular velocity for propagation to create training data is randomly chosen and set to $\bm{\omega}_0 = [0.1, -0.1, -0.25]^T~\si{deg/s}$.
The calculations in Eqs.~\eqref{eq:quaternion_omega} and \eqref{eq:euler_equation} are propagated numerically in increments of 1 s from 0 to $t_\mathrm{span, past}$ s.
In addition, the time period for the averaging of the past light curves is changed for each simulation to verify the sensitivity for estimation results.


The initial conditions for propagating the testing data, that is, the true attitude, are $\bm{q}_0 = [0, 0, 0, 1]^T$ and $\bm{\omega}_0 = [0.1, -0.1, -0.25]^{T}$ deg/s, and the observations are sampled at a frequency of 1Hz.
The surface parameters for the testing are the same as those in Table~\ref{tab:surface_para}, but the diffuse reflectance $\rho$ is changed for each simulation.
Furthermore, the parameters for the UKF are set to $(\alpha, \beta, \lambda, \kappa) = (0.0001, 2, 0, 0)$, and estimated system noise and observation noise of the UKF are set to 0.15 and 0 respectively. The hyperparameters of the GPR are set to $(\theta_1, \theta_2, \sigma) = (1, 0.25, 0.1)$.
These parameters are set heuristically by comparing the light curves with the trained Gaussian process regression and the ones with the parametric light curve model.

\begin{table}[tbp]
    \centering
    \caption{Surface Parameters.}
    \begin{tabular}{lc}
        \hline  \hline
        Parameter & Value \\
        \hline
        Reflection properties, $(n_u, n_v)$ & $(800, 800)$ \\
        Fresnel reflection, $F_0$ & 0.5 \\
        Diffuse reflectance, $\rho$ & 0.1, 0.3, 0.5, 0.7, 0.9 \\
        \hline  \hline
    \end{tabular}
    \label{tab:surface_para}
\end{table}

\subsection{Results} 
\begin{figure}[tbp]
    \centering
    \begin{minipage}[b]{0.4\linewidth}
        \centering
        \includegraphics[width=\linewidth]{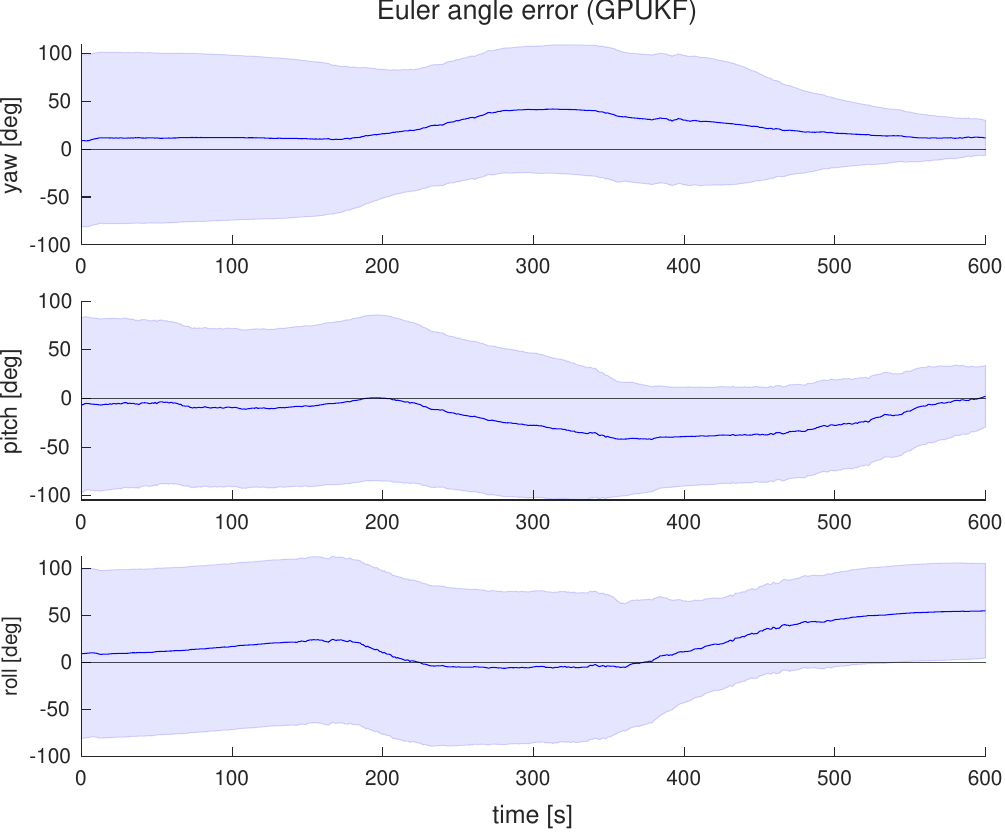}
        \subcaption{Error of estimated attitude ($t_\mathrm{span, past} = 10$).}
        \label{fig:GPUKF_rho05_tSpanPast10_eulerAngleError}
    \end{minipage}
    \hfill
    \begin{minipage}[b]{0.38\linewidth}
        \centering
        \includegraphics[width=\linewidth]{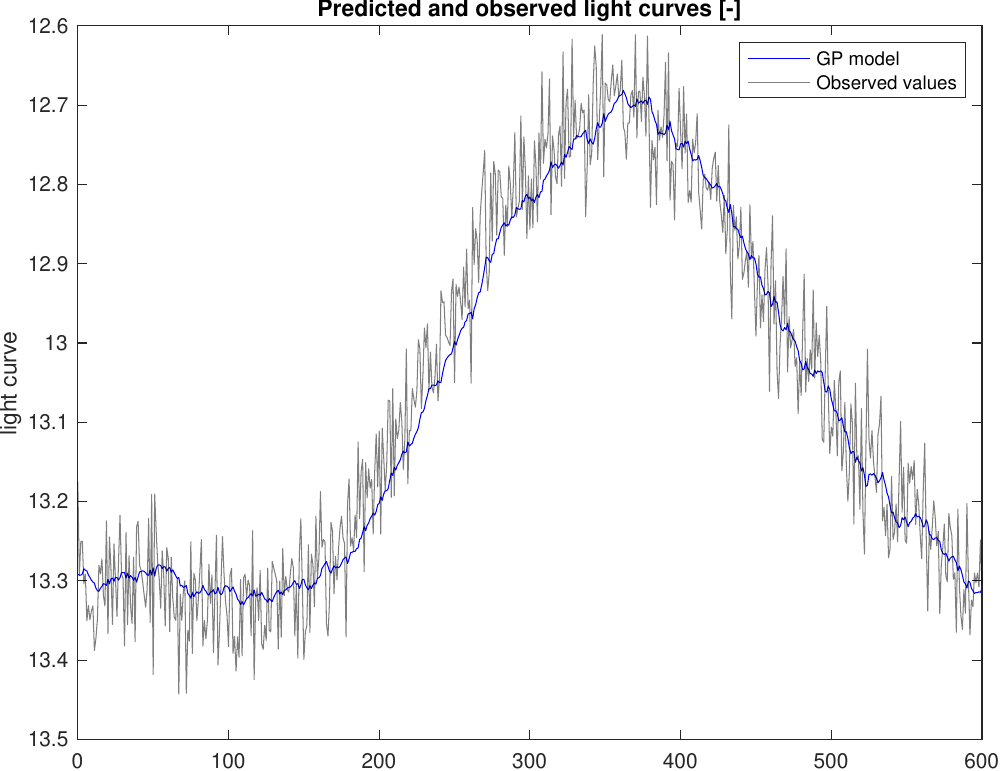}
        \subcaption{Predicted light curves.}
        \label{fig:GPUKF_rho05_tSpanPast10_observation}
    \end{minipage} \\

    \begin{minipage}[b]{0.4\linewidth}
        \centering
        \includegraphics[width=\linewidth]{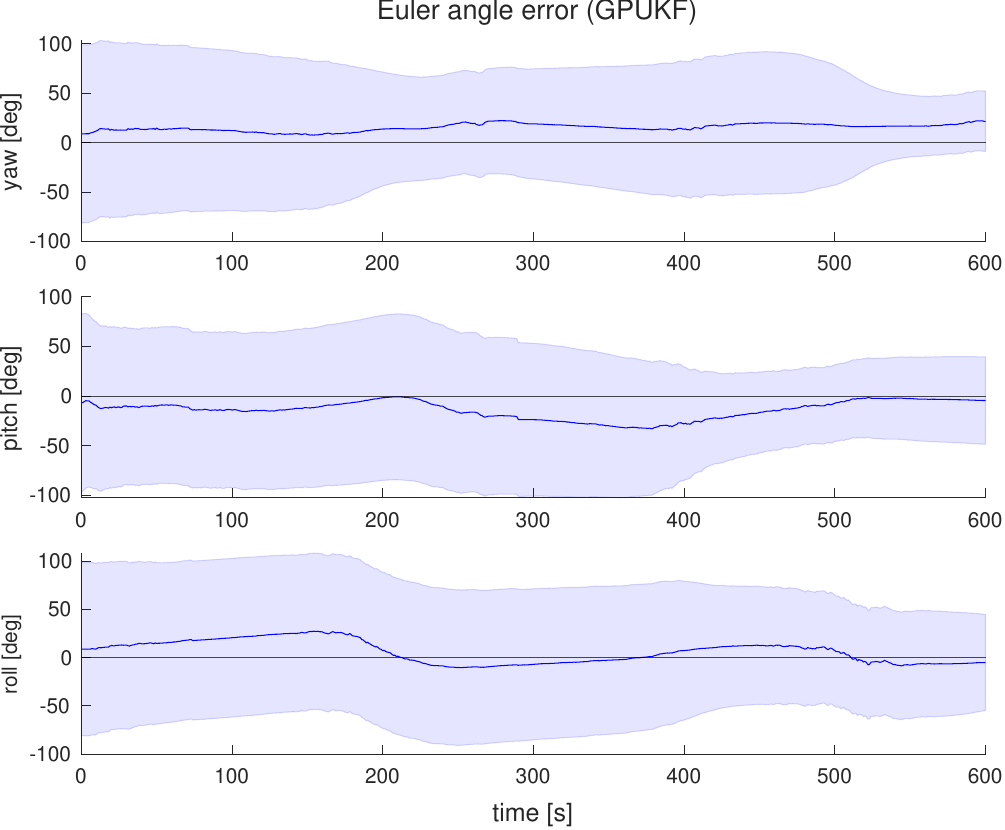}
        \subcaption{Error of estimated attitude ($t_\mathrm{span, past} = 20$).}
        \label{fig:GPUKF_rho05_tSpanPast20_eulerAngleError}
    \end{minipage}
    \hfill
    \begin{minipage}[b]{0.38\linewidth}
        \centering
        \includegraphics[width=\linewidth]{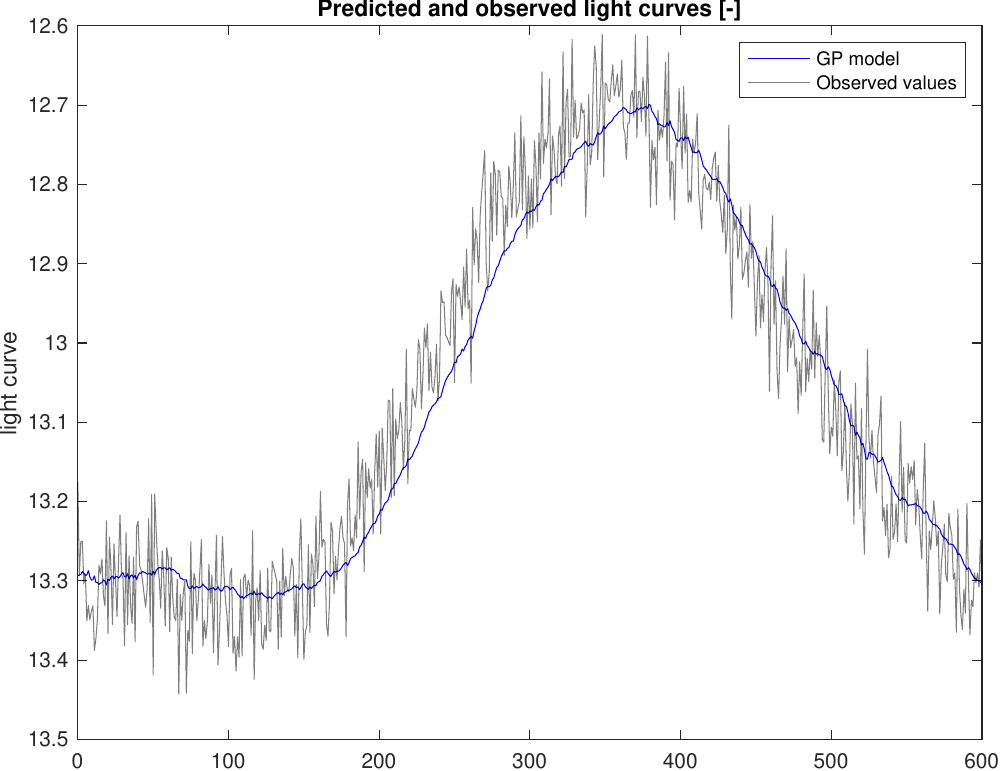}
        \subcaption{Predicted light curves.}
        \label{fig:GPUKF_rho05_tSpanPast20_observation}
    \end{minipage} \\

    
    \begin{minipage}[b]{0.4\linewidth}
        \centering
        \includegraphics[width=\linewidth]{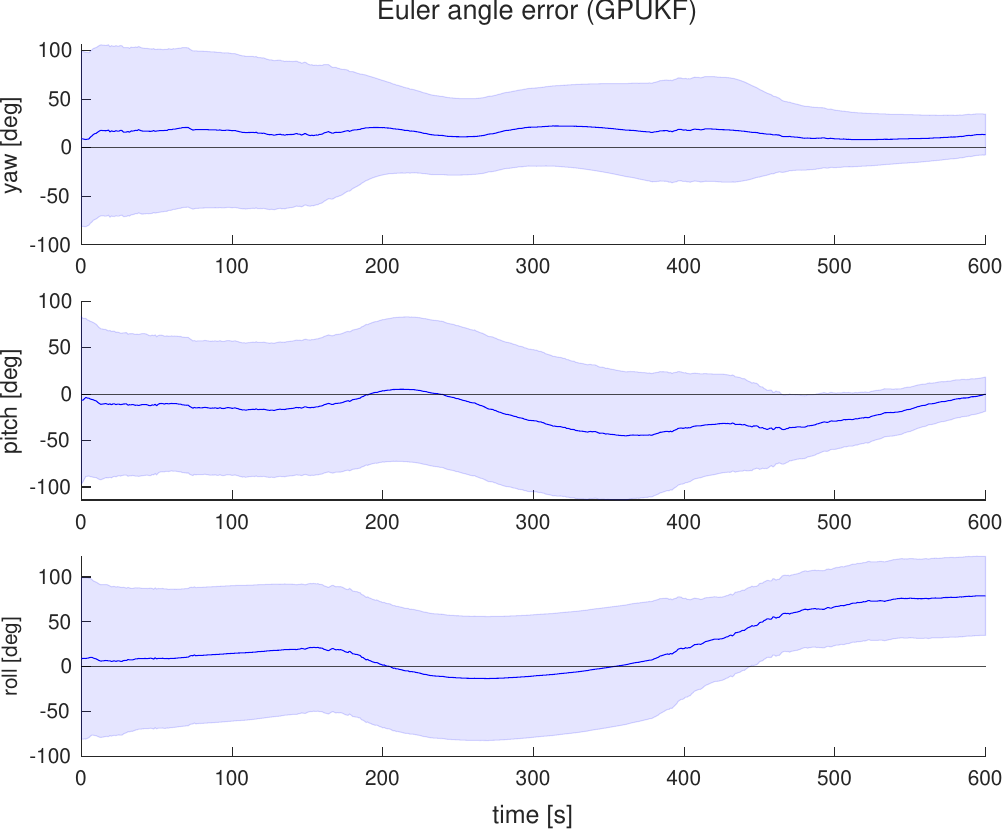}
        \subcaption{Error of estimated attitude ($t_\mathrm{span, past} = 30$).}
        \label{fig:GPUKF_rho05_tSpanPast30_eulerAngleError}
    \end{minipage}
    \hfill
    \begin{minipage}[b]{0.38\linewidth}
        \centering
        \includegraphics[width=\linewidth]{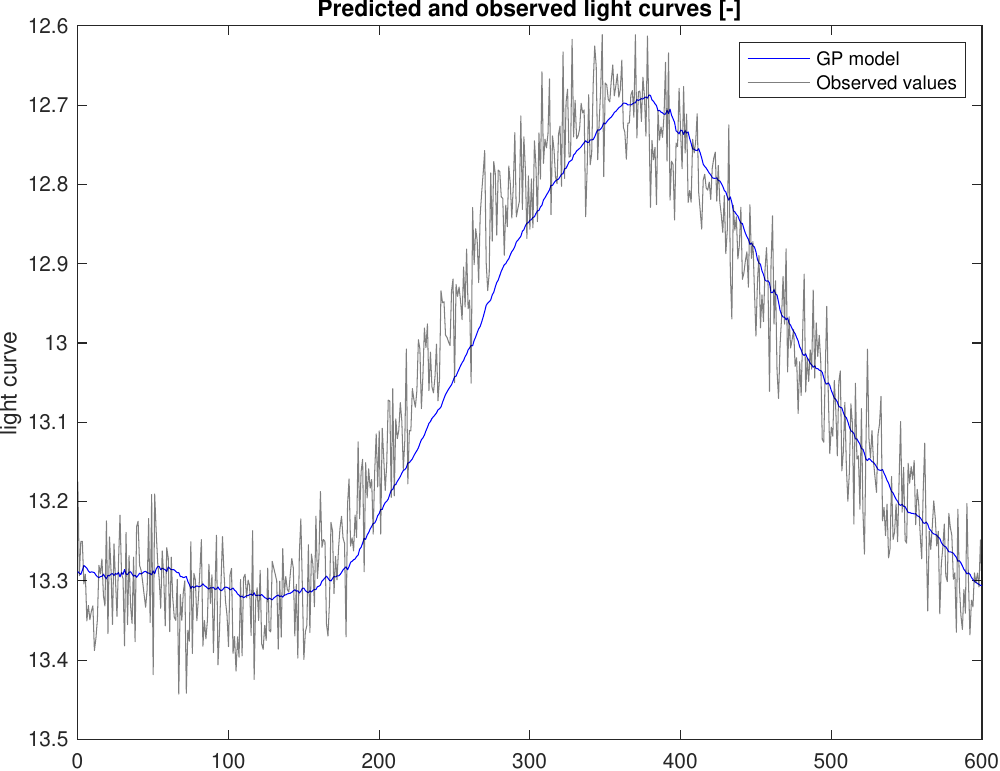}
        \subcaption{Predicted light curves.}
        \label{fig:GPUKF_rho05_tSpanPast30_observation}
    \end{minipage} \\

    \begin{minipage}[b]{0.4\linewidth}
        \centering
        \includegraphics[width=\linewidth]{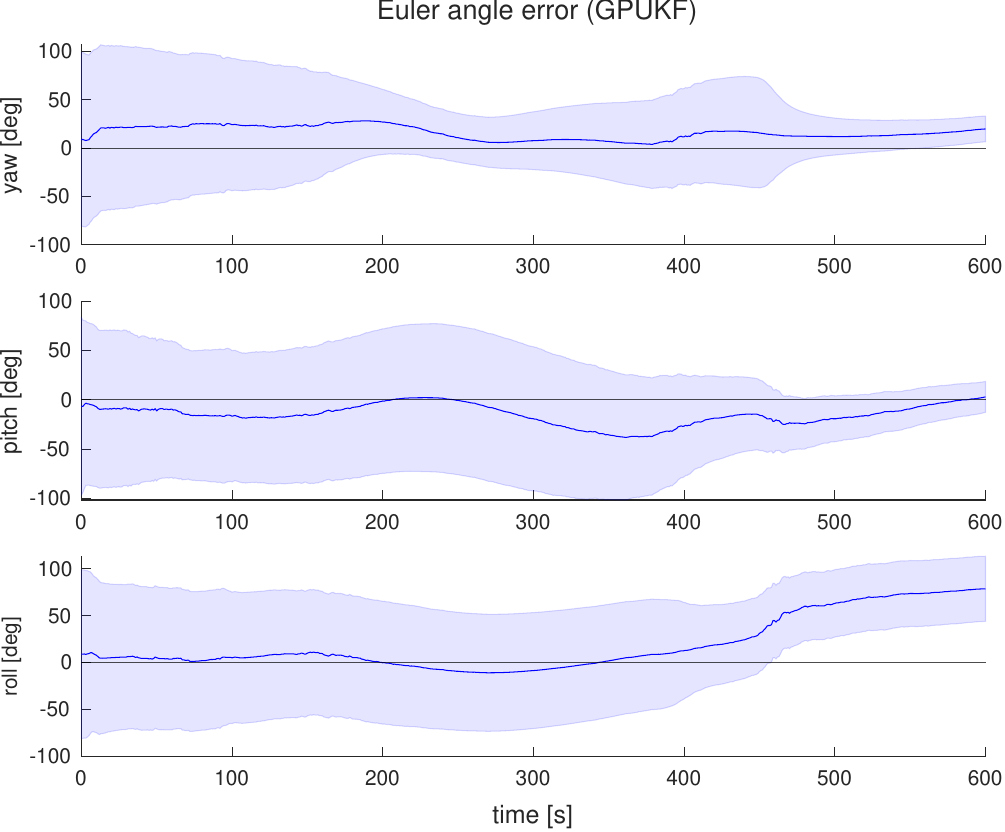}
        \subcaption{Error of estimated attitude ($t_\mathrm{span, past} = 40$).}
        \label{fig:GPUKF_rho05_tSpanPast40_eulerAngleError}
    \end{minipage}
    \hfill
    \begin{minipage}[b]{0.38\linewidth}
        \centering
        \includegraphics[width=\linewidth]{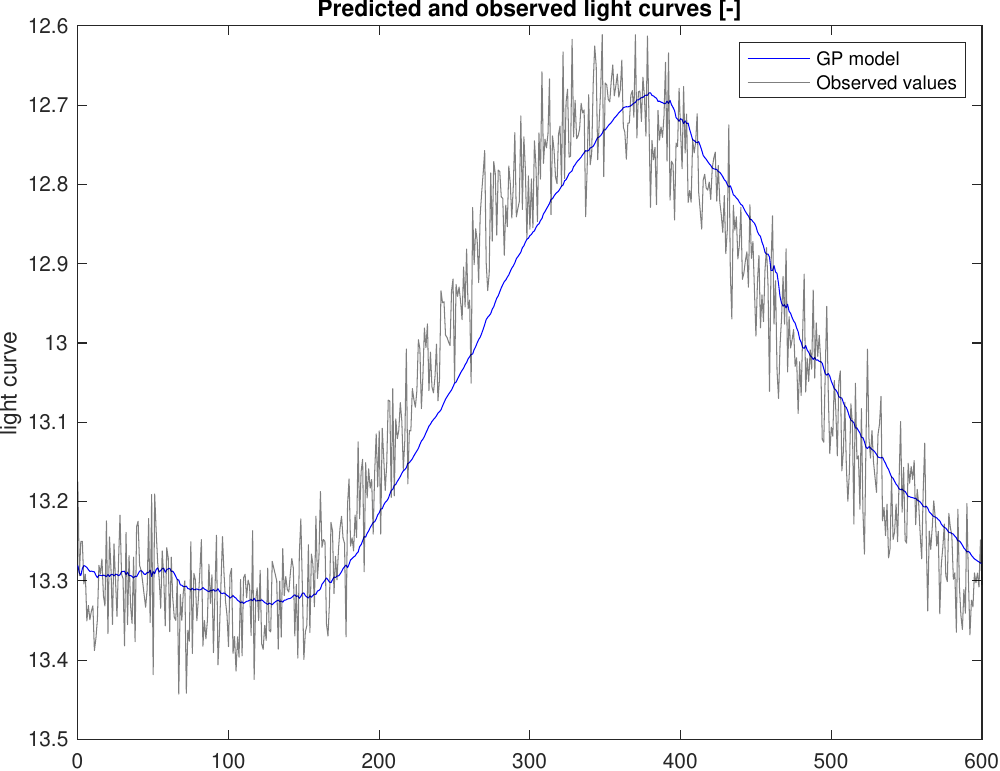}
        \subcaption{Predicted light curves.}
        \label{fig:GPUKF_rho05_tSpanPast40_observation}
    \end{minipage} \\
    
    \caption{Sensitivity analysis of the GPUKF for $t_\mathrm{span, past}$ $(\rho_\mathrm{ukf} = 0.5, \rho_\mathrm{true} = 0.5, t_\mathrm{span, past}=10, 20, 30, 40)$.}
    \label{fig:GPUKF_rho05_tSpanPast_SA}
\end{figure}

A sensitivity analysis of attitude estimation with the GPUKF is performed for various $t_\mathrm{span, past} = (10, 20, 30, 40)~{\rm s}$. The initial estimate has error $(\phi, \theta, \psi) = (-10, 5, -10)~{\rm deg}$ compared to the true value, and the observation noise of the light curves is the Gaussian distribution with a mean of 0 and a variance of 0.5.
Fig.~\ref{fig:GPUKF_rho05_tSpanPast_SA} shows the error of the attitude estimate and the predicted light curves, which are placed in the left column and the right column, respectively.
In the figures on the left, the blue line represents the estimation result of the GPUKF.
In the figures on the right, the blue line represents the predicted light curves by GPUKF and the black line is the true light curves including observation noise.
Since the input $m_\mathrm{app, past}$ in the training data is a mean of the observed light curves from the past $t_\mathrm{span, past}$, the averaged past light curves can mitigate the influence of observation noise and infer the surface parameters of the space object. 
The predicted light curves as shown in Figs.~\ref{fig:GPUKF_rho05_tSpanPast10_observation} and \ref{fig:GPUKF_rho05_tSpanPast20_observation} are better than those in Figs.~\ref{fig:GPUKF_rho05_tSpanPast30_observation} and \ref{fig:GPUKF_rho05_tSpanPast40_observation}.
Furthermore, as shown in Fig.~\ref{fig:GPUKF_rho05_tSpanPast20_eulerAngleError}, setting $t_\mathrm{span, past} =$ 20~s provides better estimation accuracy.
These results demonstrate that $t_\mathrm{span, past}=30$ and $40$ are too long to infer the surface property as $m_\mathrm{app, past}$ is updated with the observation and $t_\mathrm{span, past}=10$ is short to mitigate the influence of the observation noise.
Thus, $t_\mathrm{span, past} = 20$ is used in the following simulations.

\begin{figure}[tbp]
    \centering
    \begin{minipage}[b]{0.5\linewidth}
        \centering
        \includegraphics[width=\linewidth]{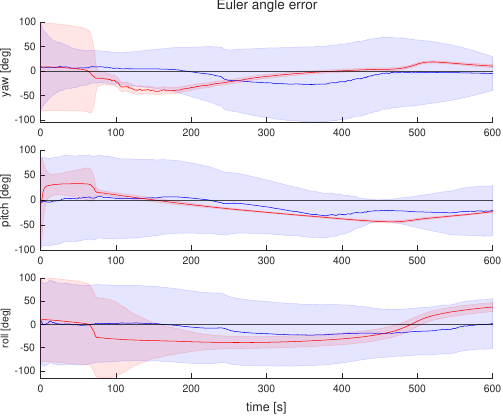}
        \subcaption{Error of estimated attitude.}
        \label{fig:cmp_rho07_eulerAngleError}
    \end{minipage}
    \hfill
    \begin{minipage}[b]{0.43\linewidth}
        \centering
        \includegraphics[width=\linewidth]{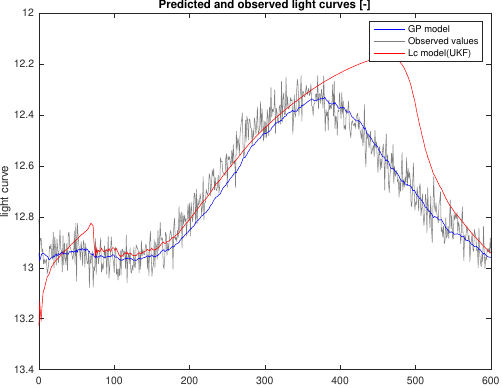}
        \subcaption{Predicted light curves.}
        \label{fig:cmp_rho07_observation}
    \end{minipage}
    \caption{Comparison of attitutde estimation using the GPUKF and the UKF $(\rho_\mathrm{ukf} = 0.5, \rho_\mathrm{true} = 0.7, t_\mathrm{span, past}=20)$.}
    \label{fig:cmp_rho07_tSpanPast20}
\end{figure}

\begin{figure}[tbp]
    \centering
    \begin{minipage}[b]{0.5\linewidth}
        \centering
        \includegraphics[width=\linewidth]{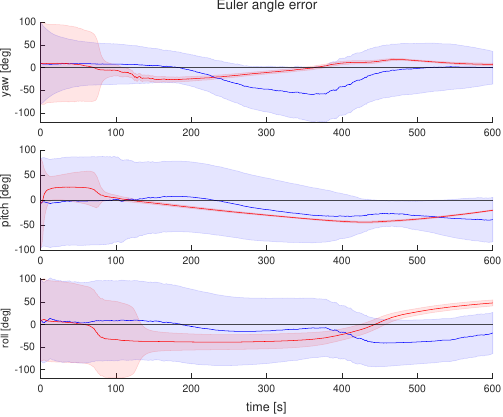}
        \subcaption{Error of estimated attitude.}
        \label{fig:cmp_rho06_eulerAngleError}
    \end{minipage}
    \hfill
    \begin{minipage}[b]{0.43\linewidth}
        \centering
        \includegraphics[width=\linewidth]{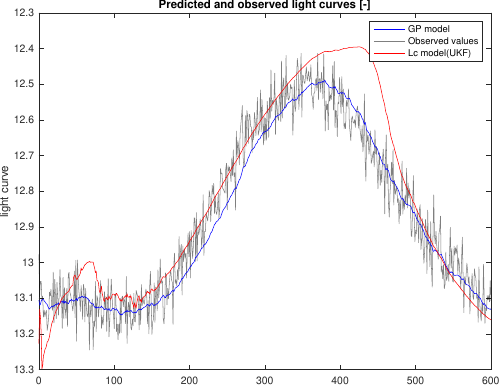}
        \subcaption{Predicted light curves.}
        \label{fig:cmp_rho06_observation}
    \end{minipage}
    \caption{Comparison of attitutde estimation using the GPUKF and the UKF $(\rho_\mathrm{ukf} = 0.5, \rho_\mathrm{true} = 0.6, t_\mathrm{span, past}=20)$.}
    \label{fig:cmp_rho06_tSpanPast20}
\end{figure}

\begin{figure}[tbp]
    \centering
    \begin{minipage}[b]{0.5\linewidth}
        \centering
        \includegraphics[width=\linewidth]{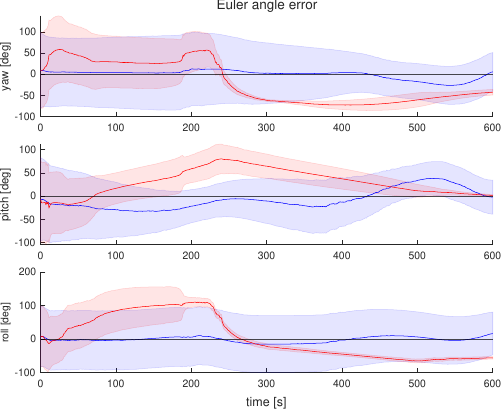}
        \subcaption{Error of estimated attitude.}
        \label{fig:cmp_rho04_eulerAngleError}
    \end{minipage}
    \hfill
    \begin{minipage}[b]{0.43\linewidth}
        \centering
        \includegraphics[width=\linewidth]{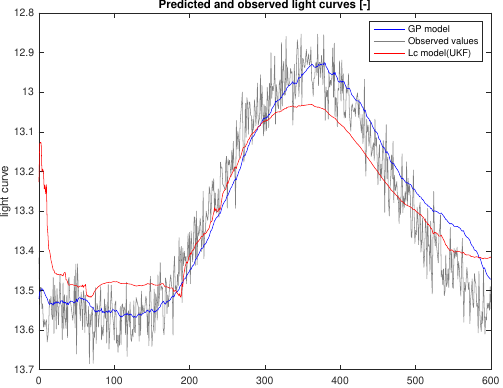}
        \subcaption{Predicted light curves.}
        \label{fig:cmp_rho04_observation}
    \end{minipage}
    \caption{Comparison of attitutde estimation using the GPUKF and the UKF $(\rho_\mathrm{ukf} = 0.5, \rho_\mathrm{true} = 0.4, t_\mathrm{span, past}=20)$.}
    \label{fig:cmp_rho04_tSpanPast20}
\end{figure}

\begin{figure}[tbp]
    \centering
    \begin{minipage}[b]{0.5\linewidth}
        \centering
        \includegraphics[width=\linewidth]{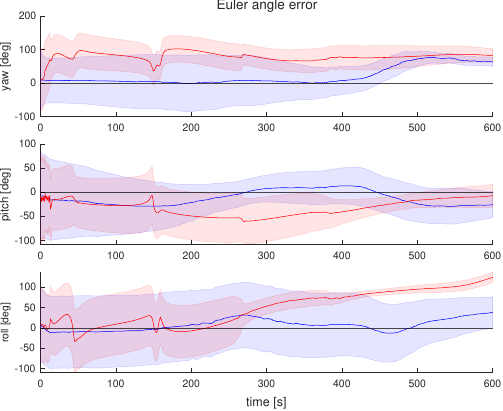}
        \subcaption{Error of estimated attitude.}
        \label{fig:cmp_rho03_eulerAngleError}
    \end{minipage}
    \hfill
    \begin{minipage}[b]{0.43\linewidth}
        \centering
        \includegraphics[width=\linewidth]{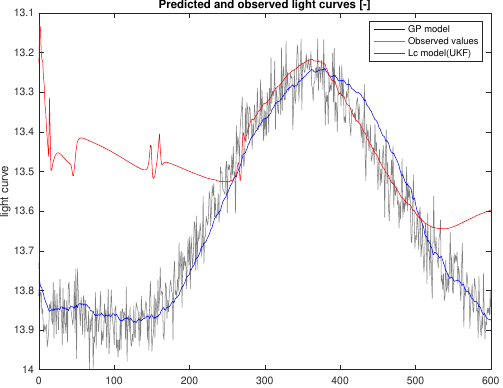}
        \subcaption{Predicted light curves.}
        \label{fig:cmp_rho03_observation}
    \end{minipage}
    \caption{Comparison of attitutde estimation using the GPUKF and the UKF $(\rho_\mathrm{ukf} = 0.5, \rho_\mathrm{true} = 0.3, t_\mathrm{span, past}=20)$.}
    \label{fig:cmp_rho03_tSpanPast20}
\end{figure}

The robustness against uncertainties in surface properties of the proposed method is examined for four cases as shown in Figs.~\ref{fig:cmp_rho07_tSpanPast20}--\ref{fig:cmp_rho03_tSpanPast20}, where the target object has the diffuse reflectance $\rho_\mathrm{true}=(0.7, 0.6, 0.4, 0.3)$.
In the same way as the above sensitivity analysis, the initial estimate has error $(\phi, \theta, \psi) = (-10, 5, -10)~{\rm deg}$ compared to the true value, and the observation noise of the light curves is the Gaussian distribution with a mean of 0 and a variance of 0.5.
The figures on the left show the errors of attitude estimate, where the blue line indicates the GPUKF and the red line indicates the conventional UKF.
The figures on the right show the predicted light curves, where the blue line represents the ones by GPUKF, the red line represents the ones by the conventional UKF, and the black line represents the true light curves including observation noise.
The calculation time for an observation step is summarized in Table~\ref{tab:computational_cost}.
The approximate entire computation times of the proposed method (GPUKF) and the conventional method (UKF) for the estimation are 9.0 and 0.2 minutes respectively.
\begin{table}[tbp]
    \centering
    \caption{Computational cost for an observation step.}
    \begin{tabular}{lcc}
        \hline  \hline
        Result & Calculation time (GPUKF) & Calculation time (UKF) \\
        \hline
        Fig.~\ref{fig:cmp_rho07_tSpanPast20}& $0.8793~\si{s}$ & $0.006887~\si{s}$ \\
        Fig.~\ref{fig:cmp_rho06_tSpanPast20} & $0.8476~\si{s}$ & $0.005964~\si{s}$\\

        Fig.~\ref{fig:cmp_rho04_tSpanPast20} & $0.8289~\si{s}$ & $0.002959~\si{s}$ \\

        Fig.~\ref{fig:cmp_rho03_tSpanPast20} & $0.8382~\si{s}$ & $0.005509~\si{s}$ \\
        
        \hline  \hline
    \end{tabular}
    \label{tab:computational_cost}
\end{table}


        
In all four cases shown in Figs.~\ref{fig:cmp_rho07_tSpanPast20}--\ref{fig:cmp_rho03_tSpanPast20}, the attitude estimate and the predicted light curve by GPUKF have better or comparable accuracy than those by the conventional UKF. 
When the diffuse reflectance $\rho$ used in the conventional UKF is different from the true value (i.e., $\rho$ is unknown), the conventional UKF cannot predict the light curves and estimate the attitude as shown in Fig.~\ref{fig:ukf_doesNotWork_results}.
On the other hand, the GPUKF can predict the light curves and estimate the attitude better than the conventional UKF as shown in Figs.~\ref{fig:cmp_rho07_tSpanPast20}--\ref{fig:cmp_rho03_tSpanPast20}, demonstrating that the GPUKF method has the robustness against uncertainties in surface properties.
In fact, the Root mean squared error (RMSE) of each result of the GPUKF method is obviously better than the conventional UKF method as summarized in Table~\ref{tab:rmse_attitude} and \ref{tab:rmse_lightcurve}.
These RMSE are calculated for the estimated values from 400~s to 600~s.

\begin{table}[tbp]
    \centering
    \caption{Root mean squared error of estimated attitude (yaw, pitch, roll).}
    \begin{tabular}{lcc}
        \hline  \hline
        Result & RMSE (GPUKF) & RMSE (UKF) \\
        \hline
        \small
        Fig.~\ref{fig:cmp_rho07_eulerAngleError}& $(8.1251, 23.7262, 14.4020)~\si{deg}$ & $(11.5760, 37.0181, 24.6825)~\si{deg}$\\
        Fig.~\ref{fig:cmp_rho06_eulerAngleError} & $(14.5384, 33.3480, 34.0134)~\si{deg}$ & $(12.4030, 37.0327, 30.2688)~\si{deg}$\\

        Fig.~\ref{fig:cmp_rho04_eulerAngleError} & $(15.5913, 23.2875, 6.8249)~\si{deg}$ & $(60.7433, 18.6831, 57.4037)~\si{deg}$ \\

        Fig.~\ref{fig:cmp_rho03_eulerAngleError} & $(57.5681, 21.5917, 19.4448)~\si{deg}$ & $(80.0133, 22.2466, 96.0530)~\si{deg}$ \normalsize \\     
        \hline  \hline
    \end{tabular}
    \label{tab:rmse_attitude}
\end{table}

\begin{table}[tbp]
    \centering
    \caption{Root mean squared error of predicted light curve.}
    \begin{tabular}{lcc}
        \hline  \hline
        Result & RMSE (GPUKF) & RMSE (UKF) \\
        \hline
        Fig.~\ref{fig:cmp_rho07_observation}& 0.9978 & 15.3630 \\
        Fig.~\ref{fig:cmp_rho06_observation} & 1.3885 & 7.5295 \\

        Fig.~\ref{fig:cmp_rho04_observation} & 4.8400 & 3.7277 \\

        Fig.~\ref{fig:cmp_rho03_observation} & 3.0916 & 6.4077 \\
        
        \hline  \hline
    \end{tabular}
    \label{tab:rmse_lightcurve}
\end{table}


        



        

\section{Conclusions}
This paper addresses a non-parametric model of light curves using Gaussian process regression (GPR).
When the attitude estimation via light curves is conducted for the resident space object (RSO) with unknown surface properties, the conventional estimation scheme suffers from the difficulty that the observation model cannot be constructed parametrically because the light curves depend on surface properties.
This study proposes an estimation method to describe the observation model using GP and combines it with UKF (GPUKF) for the estimation of the attitude using light curves.
The training data set of GPUKF is generated for various surface parameters so that robustness against the uncertainty of the target object's surface property is obtained.
Numerical simulations show that the attitude estimation using GPUKF has a better accuracy than that using a conventional UKF.
Furthermore, the comparison between GPUKF and UKF shows that the GPUKF can regress the light curves under conditions that the UKF cannot. 
The attitude estimation of the non-resolved object via light curves can be more developed by refining the proposed method, contributing to practical scenarios in SSA/SDA where object parameters are unknown.



\bibliographystyle{elsarticle-num} 
\bibliography{ref}

@article{cowardin2024orbital,
  title={Orbital Debris Quarterly News, July 2024},
  author={Cowardin, Heather},
  journal={Orbital Debris Quarterly News},
  volume={28},
  number={3},
  year={2024},
  publisher={National Aeronautics and Space Administration}
}

@article{curzi2020large,
  title={Large constellations of small satellites: A survey of near future challenges and missions},
  author={Curzi, Giacomo and Modenini, Dario and Tortora, Paolo},
  journal={Aerospace},
  volume={7},
  number={9},
  pages={133},
  year={2020},
  publisher={MDPI}
}

@article{kaasalainen2001optimization,
  title={Optimization methods for asteroid lightcurve inversion: I. shape determination},
  author={Kaasalainen, Mikko and Torppa, Johanna},
  journal={Icarus},
  volume={153},
  number={1},
  pages={24--36},
  year={2001},
  publisher={Elsevier}
}

@article{muinonen2020asteroid,
  title={Asteroid lightcurve inversion with Bayesian inference},
  author={Muinonen, K and Torppa, J and Wang, X-B and Cellino, A and Penttil{\"a}, A},
  journal={Astronomy \& Astrophysics},
  volume={642},
  pages={A138},
  year={2020},
  publisher={EDP sciences}
}

@Article{matsu,
author = {MATSUSHITA, Yuri and YOSHIMURA, Yasuhiro and NAGASAKI, Shuji and HANADA, Toshiya}, 
title = {Conceptual Study of Improved Photometric Attitude Estimation Using Glint}, 
journal = {TRANSACTIONS OF THE JAPAN SOCIETY FOR AERONAUTICAL AND SPACE SCIENCES, AEROSPACE TECHNOLOGY JAPAN}, 
volume = {22}, 
number = {0}, 
pages = {59–65}, 
year = {2024}, }

@article{muller2017hayabusa,
  title={Hayabusa-2 mission target asteroid 162173 Ryugu (1999 JU3): Searching for the object’s spin-axis orientation},
  author={M{\"u}ller, TG and {\v{D}}urech, J and Ishiguro, Masateru and Mueller, Michael and Kr{\"u}hler, T and Yang, Hao and Kim, M-J and O’Rourke, Laurence and Usui, Fumihiko and Kiss, Casba and others},
  journal={Astronomy \& Astrophysics},
  volume={599},
  pages={A103},
  year={2017},
  publisher={EDP Sciences}
}

@book{books/lib/RasmussenW06,
  author = {Rasmussen, Carl Edward and Williams, Christopher K. I.},
  ee = {https://www.worldcat.org/oclc/61285753},
  isbn = {026218253X},
  pages = {I-XVIII, 1-248},
  publisher = {MIT Press},
  series = {Adaptive computation and machine learning},
  title = {Gaussian processes for machine learning.},
  year = 2006
}

@ARTICLE{7819454,
  author={Piergentili, Fabrizio and Santoni, Fabio and Seitzer, Patrick},
  journal={IEEE Transactions on Aerospace and Electronic Systems}, 
  title={Attitude Determination of Orbiting Objects from Lightcurve Measurements}, 
  year={2017},
  volume={53},
  number={1},
  pages={81-90},
  doi={10.1109/TAES.2017.2649240}}

@inproceedings{burton2023fast,
  title={Fast Light Curve Inversion for Regular and Tumbling Attitude Motion},
  author={Burton, Alexander and Frueh, Carolin},
  booktitle={Proceedings of the Advanced Maui Optical and Space Surveillance (AMOS) Technologies Conference},
  pages={83},
  year={2023}
}

@article{kwast2014introduction,
  title={An introduction to brdf models},
  author={Kwast, Dani{\"e}l Jimenez},
  journal={Hmi. Ewi. Utwente, NI},
  year={2014},
  publisher={Citeseer}
}

@article{wetterer2009attitude,
  title={Attitude determination from light curves},
  author={Wetterer, Charles J and Jah, Moriba},
  journal={Journal of Guidance, Control, and Dynamics},
  volume={32},
  number={5},
  pages={1648--1651},
  year={2009}
}

@article{ashikhmin2000anisotropic,
  title={An anisotropic phong brdf model},
  author={Ashikhmin, Michael and Shirley, Peter},
  journal={Journal of graphics tools},
  volume={5},
  number={2},
  pages={25--32},
  year={2000},
  publisher={Taylor \& Francis}
}

@article{crassidis2003unscented,
  title={Unscented filtering for spacecraft attitude estimation},
  author={Crassidis, John L and Markley, F Landis},
  journal={Journal of guidance, control, and dynamics},
  volume={26},
  number={4},
  pages={536--542},
  year={2003}
}

@inproceedings{wan2000unscented,
  title={The unscented Kalman filter for nonlinear estimation},
  author={Wan, Eric A and Van Der Merwe, Rudolph},
  booktitle={Proceedings of the IEEE 2000 adaptive systems for signal processing, communications, and control symposium (Cat. No. 00EX373)},
  pages={153--158},
  year={2000},
  organization={Ieee}
}

@article{du2018attitude,
  title={The attitude inversion method of geostationary satellites based on unscented particle filter},
  author={Du, Xiaoping and Wang, Yang and Hu, Heng and Gou, Ruixin and Liu, Hao},
  journal={Advances in Space Research},
  volume={61},
  number={8},
  pages={1984--1996},
  year={2018},
  publisher={Elsevier}
}

@article{cabrera2023adaptive,
  title={The adaptive Gaussian mixtures unscented Kalman filter for attitude determination using light curves},
  author={Cabrera, David Vallverd{\'u} and Utzmann, Jens and F{\"o}rstner, Roger},
  journal={Advances in Space Research},
  volume={71},
  number={6},
  pages={2609--2628},
  year={2023},
  publisher={Elsevier}
}

@book{10.5555/1121596,
author = {Thrun, Sebastian and Burgard, Wolfram and Fox, Dieter},
title = {Probabilistic Robotics (Intelligent Robotics and Autonomous Agents)},
year = {2005},
isbn = {0262201623},
publisher = {The MIT Press}
}





\end{document}